\documentclass[a4paper,11pt]{article}
\pdfoutput=1 

\usepackage{jheppub} 

\usepackage[T1]{fontenc} 

\usepackage{enumerate}
\usepackage{graphicx}
\usepackage{color}
\usepackage{physics}
\usepackage{url,hyperref}
\usepackage{mathrsfs}
\usepackage{amssymb}
\usepackage{amsmath}
\usepackage{amsthm}
\usepackage[mathscr]{euscript}
\usepackage{subcaption} 
\usepackage{bm}
\DeclareMathAlphabet\mathbfcal{OMS}{cmsy}{b}{n}
\usepackage[dvipsnames]{xcolor}

\usepackage{tikz}
\usetikzlibrary{positioning, calc}
\usetikzlibrary{calc}
\usetikzlibrary{arrows}
\usepackage{tikz-3dplot}
\usetikzlibrary{fadings}
\usetikzlibrary{decorations.pathreplacing,decorations.markings,decorations.pathmorphing}
\tikzset{snake it/.style={decorate, decoration=snake}}
\usetikzlibrary{patterns,patterns.meta}
\usetikzlibrary{decorations}

\tikzset{
    >=stealth',
    punkt/.style={
           rectangle,
           rounded corners,
           draw=black, very thick,
           text width=6.5em,
           minimum height=2em,
           text centered},
    pil/.style={
           ->,
           thick,
           shorten <=2pt,
           shorten >=2pt,},
  on each segment/.style={
    decorate,
    decoration={
      show path construction,
      moveto code={},
      lineto code={
        \path [#1]
        (\tikzinputsegmentfirst) -- (\tikzinputsegmentlast);
      },
      curveto code={
        \path [#1] (\tikzinputsegmentfirst)
        .. controls
        (\tikzinputsegmentsupporta) and (\tikzinputsegmentsupportb)
        ..
        (\tikzinputsegmentlast);
      },
      closepath code={
        \path [#1]
        (\tikzinputsegmentfirst) -- (\tikzinputsegmentlast);
      },
    },
  },
  mid arrow/.style={postaction={decorate,decoration={
        markings,
        mark=at position .5 with {\arrow[#1]{stealth'}}
      }}}
}

\newtheorem{theorem}{Theorem}

\theoremstyle{definition}

\newtheorem{definition}[theorem]{Definition}

\title{Non-local computation and the black hole interior}

\author[a,b]{Alex May,}
\author[a]{Michelle Xu.}

\affiliation[a]{Stanford Institute for Theoretical Physics, Stanford University, 382 Via Pueblo Mall, Stanford, CA 94305-4060, U.S.A.}
\affiliation[b]{Perimeter Institute for Theoretical Physics, Waterloo, Ontario N2L 2Y5, Canada}

\emailAdd{amay@perimeterinstitute.ca}
\emailAdd{mdx@stanford.edu}

\abstract{
In a two sided black hole, systems falling in from opposite asymptotic regions can meet inside the black hole and interact. 
This is the case even while the two CFTs describing each asymptotic region are non-interacting. 
Here, we relate these behind the horizon interactions to non-local quantum computations. 
This gives a quantum circuit perspective on these interactions, which applies whenever the interaction occurs in the past of a certain extremal surface that sits inside the black hole and in arbitrary dimension. 
Whenever our perspective applies, we obtain a boundary signature for these interior collisions which is stated in terms of the mutual information. 
We further revisit the connection discussed earlier between bulk interactions in one sided AdS geometries and non-local computation, and recycle some of our techniques to offer a new perspective on making that connection precise. 
}

\begin{document} 
\maketitle
\flushbottom

\section{Introduction}\label{sec:intro}

In the AdS/CFT correspondence, local bulk interactions are supported by local interactions in a lower dimensional boundary description. 
How this can be possible is puzzling, in a way that is most sharp in the context of two sided black hole solutions. 
There, excitations originating in each of the two asymptotic regions can meet and interact behind the black hole horizon. 
For instance, we can imagine an observer, Alice, who is created by acting on the left CFT, and a second observer Bob who is created by acting on the right CFT. 
While the CFTs don't interact, the bulk description of their time evolution involves Alice and Bob meeting and interacting behind the horizon. 
After being initially emphasized by Marolf and Wall \cite{marolf2012eternal}, how to understand these behind the horizon interactions as emerging from a boundary perspective has remained puzzling, and many perspectives on this have been offered \cite{de2022black,gao2022seeing,jafferis2022inside,zhao2021collision,haehl2021diagnosing,haehl2021six,haehl2022collisions}. 

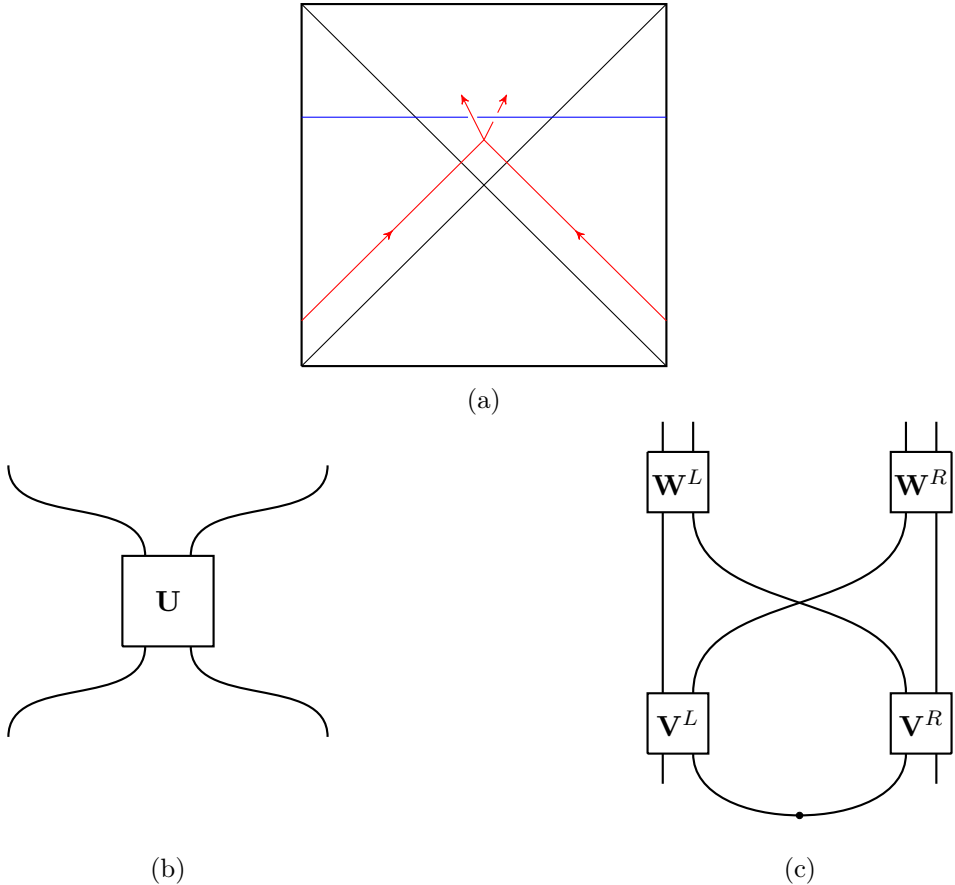
\begin{figure*}
    \centering
    \begin{subfigure}{0.55\textwidth}
    \centering
    \begin{tikzpicture}[scale=0.6]
    
    \draw[thick] (-4,-4) -- (-4,4) -- (4,4) -- (4,-4) -- (-4,-4);
    \draw (-4,-4) -- (4,4);
    \draw (-4,4) -- (4,-4);

    \draw[red,postaction={on each segment={mid arrow}}] (-4,-3) -- (0,1);
    \draw[red,postaction={on each segment={mid arrow}}] (4,-3) -- (0,1);
    \draw[red] (0,1) -- (0.2,1.4);
    \draw[red,->] (0.3,1.6) -- (0.5,2);
    \draw[blue] (-4,1.5) -- (-0.35,1.5);
    \draw[blue] (-0.15,1.5) -- (4,1.5);
    \draw[red,->] (0,1) -- (-0.5,2);
    
    \end{tikzpicture}
    \caption{}
    \label{fig:intropenrose}
    \end{subfigure}
    \hfill
    \centering
    \begin{subfigure}{0.45\textwidth}
    \centering
    \begin{tikzpicture}[scale=0.6]
    
    \draw[thick] (-1,-1) -- (-1,1) -- (1,1) -- (1,-1) -- (-1,-1);
    
    \draw[thick] (-3.5,-3) to [out=90,in=-90] (-0.5,-1);
    \draw[thick] (3.5,-3) to [out=90,in=-90] (0.5,-1);
    
    \draw[thick] (0.5,1) to [out=90,in=-90] (3.5,3);
    \draw[thick] (-0.5,1) to [out=90,in=-90] (-3.5,3);
    
    \node at (0,0) {$\mathbf{U}$};

    \node at (0,-5) {$ $};
    
    \end{tikzpicture}
    \caption{}
    \label{fig:local}
    \end{subfigure}
    \hfill
    \begin{subfigure}{0.45\textwidth}
    \centering
    \begin{tikzpicture}[scale=0.4]
    
    \draw[thick] (-5,-5) -- (-5,-3) -- (-3,-3) -- (-3,-5) -- (-5,-5);
    \node at (-4,-4) {$\mathbf{V}^L$};
    
    \draw[thick] (5,-5) -- (5,-3) -- (3,-3) -- (3,-5) -- (5,-5);
    \node at (4,-4) {$\mathbf{V}^R$};
    
    \draw[thick] (5,5) -- (5,3) -- (3,3) -- (3,5) -- (5,5);
    \node at (4,4) {$\mathbf{W}^R$};
    
    \draw[thick] (-5,5) -- (-5,3) -- (-3,3) -- (-3,5) -- (-5,5);
    \node at (-4,4) {$\mathbf{W}^L$};
    
    \draw[thick] (-4.5,-3) -- (-4.5,3);
    
    \draw[thick] (4.5,-3) -- (4.5,3);
    
    \draw[thick] (-3.5,-3) to [out=90,in=-90] (3.5,3);
    
    \draw[thick] (3.5,-3) to [out=90,in=-90] (-3.5,3);
    
    \draw[thick] (-3.5,-5) to [out=-90,in=-90] (3.5,-5);
    \draw[black] plot [mark=*, mark size=3] coordinates{(0,-7.05)};
    
    \draw[thick] (-4.5,-6) -- (-4.5,-5);
    \draw[thick] (4.5,-6) -- (4.5,-5);
    
    \draw[thick] (4.5,5) -- (4.5,6);
    \draw[thick] (-4.5,5) -- (-4.5,6);
    
    \draw[thick] (3.5,5) -- (3.5,6);
    \draw[thick] (-3.5,5) -- (-3.5,6);
    
    \end{tikzpicture}
    \caption{}
    \label{fig:non-localcomputation}
    \end{subfigure}
    \caption{(a) Two systems (red) fall into a black hole from opposite asymptotic regions and interact. We assume they meet before an entangling surface (blue) that threads through the hole  (b) Circuit diagram showing the local implementation of a unitary in terms of a unitary $\mathbf{U}$. The bulk AdS picture for the interacting systems takes this form. (c) Circuit diagram showing the non-local implementation of a unitary $\mathbf{U}$. We show the bulk interaction from figure (a) must be reproducible in this form, using an amount of entanglement given by the area of the black hole.}
    \label{fig:non-localandlocal}
\end{figure*}

In this article, we relate behind the horizon interactions to a circuit construction in quantum information theory known as a non-local quantum computation, illustrated in figure \ref{fig:non-localandlocal}.
Our discussion is closely related to earlier work which argued bulk interactions in spacetimes with a single asymptotically AdS boundary are supported by non-local computations in the CFT \cite{may2019quantum,may2020holographic,may2021holographic,dolev2022holography,may2022connected}.
Our main contribution is to extend this to the setting of two sided black hole geometries, where there is a tension between having local bulk interactions and two boundary theories which do not interact. 
Our construction also works in arbitrary dimensions, extending earlier results which apply to $2+1$ bulk dimensions. 

The basic setting we study is of a two sided AdS$_{d+1}$ black hole, dual to two CFTs prepared in the thermofield double state. 
Considering a CFT subregion $R$ which includes portions of both the left and right CFTs the corresponding entanglement wedge $E_{R}$, which represents the portion of the bulk which can be reconstructed from $R$, can include portions of the black hole interior. 
This happens when the relevant entangling surface threads through the black hole and connects the two asymptotic regions \cite{hartman2013time}.
We leverage this geometrical observation and entanglement wedge reconstruction to argue behind the horizon interactions can be reproduced in the form of a non-local computation, at least when they happen in the past of an entangling surface that reaches through the black hole (see figure \ref{fig:intropenrose}). 
We refer to these as `early' collisions. 
Interestingly, this last requirement means interactions that happen sufficiently deep in the black hole interior cannot be understood as non-local computations. 

More quantitatively, we are interested in understanding how much entanglement is necessary to support interaction within a given bulk region. 
In $2+1$ bulk dimensions, we study the spherical and planar BTZ black hole and study when bulk interactions can be reproduced as non-local computations. 
By adjusting the length of the transverse direction of the planar black hole, or temperature of the spherical black hole, we can reproduce interactions happening deeper within the hole. 
For pairs of CFTs sharing finite entanglement however we always find there is a hidden region within the hole, within which we cannot understand interactions in these terms. 
We leave better understanding interactions in this deep interior region to future work. 

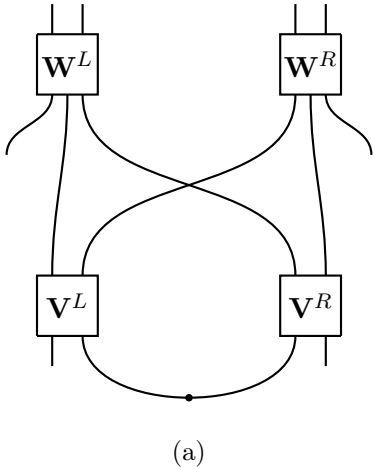
\begin{figure*}
    \centering
    \begin{subfigure}{0.45\textwidth}
    \centering
    \begin{tikzpicture}[scale=0.4]
    
    \draw[thick] (-5,-5) -- (-5,-3) -- (-3,-3) -- (-3,-5) -- (-5,-5);
    \node at (-4,-4) {$\mathbf{V}^L$};
    
    \draw[thick] (5,-5) -- (5,-3) -- (3,-3) -- (3,-5) -- (5,-5);
    \node at (4,-4) {$\mathbf{V}^R$};
    
    \draw[thick] (5,5) -- (5,3) -- (3,3) -- (3,5) -- (5,5);
    \node at (4,4) {$\mathbf{W}^R$};
    \draw[thick] (6,1) to [out=90,in=-90] (4.5,3);

    \draw[thick] (-6,1) to [out=90,in=-90] (-4.5,3);
    
    \draw[thick] (-5,5) -- (-5,3) -- (-3,3) -- (-3,5) -- (-5,5);
    \node at (-4,4) {$\mathbf{W}^L$};
    
    \draw[thick] (-4.5,-3) to [out=90,in=-90] (-4,3);
    
    \draw[thick] (4.5,-3) to [out=90,in=-90] (4,3);
    
    \draw[thick] (-3.5,-3) to [out=90,in=-90] (3.5,3);
    
    \draw[thick] (3.5,-3) to [out=90,in=-90] (-3.5,3);
    
    \draw[thick] (-3.5,-5) to [out=-90,in=-90] (3.5,-5);
    \draw[black] plot [mark=*, mark size=3] coordinates{(0,-7.05)};
    
    \draw[thick] (-4.5,-6) -- (-4.5,-5);
    \draw[thick] (4.5,-6) -- (4.5,-5);
    
    \draw[thick] (4.5,5) -- (4.5,6);
    \draw[thick] (-4.5,5) -- (-4.5,6);
    
    \draw[thick] (3.5,5) -- (3.5,6);
    \draw[thick] (-3.5,5) -- (-3.5,6);
    
    \end{tikzpicture}
    \caption{}
    \label{}
    \end{subfigure}
    \caption{For AdS geometries with a single asymptotic boundary, bulk computations are naively related to the augmented non-local computation shown here. Comparing to the (non-augmented) form of a non-local computation (figure \ref{fig:non-localcomputation}), extra systems enter the circuit in the second round.}
    \label{fig:augmentednon-localcomputation}
\end{figure*}

As a concrete byproduct of our perspective on interior interactions, we point out a boundary signature of early interior collisions. 
To find this boundary signature, we apply the connected wedge theorem of \cite{may2019quantum,may2020holographic,may2021holographic,may2022connected} to the setting of a two-sided black hole.\footnote{In fact, a slight extension of the connected wedge theorem is required: we argue for this extension heuristically but omit a detailed treatment. We expect a formal proof is possible using the techniques of \cite{may2022connected}.}
This leads to the statement that early interior collisions imply large mutual information between associated boundary regions. 

\vspace{0.2cm}
\noindent \textbf{Relationship to earlier work}
\vspace{0.2cm}

It is useful to compare our discussion to an earlier one \cite{may2019quantum,may2020holographic,may2021holographic} addressing the single sided setting. 
There, interactions in the bulk were most easily related to boundary processes with the form shown in figure \ref{fig:augmentednon-localcomputation}, which we refer to as an augmented non-local computation. 
Notice that compared to a standard non-local computation, extra quantum systems appear in the second round operations. 
While we could now choose to consider the (apparently only slightly more complicated) boundary description, these extra systems actually make the boundary circuit much more difficult to understand and constrain. 
In particular, these circuits can support interactions while having the reduced density matrix describing the state held at the two input locations be unentangled \cite{dolev2022holography}.
This obscures the role entanglement is playing in supporting local bulk interaction. 
In the two sided setting, this complication is naturally removed, and bulk interactions are directly related to the standard non-local computation scenario.

To relate interactions in the one-sided setting to non-local computation, \cite{may2019quantum,may2020holographic,may2021holographic} argued on heuristic grounds that the (non-augmented) non-local computation was the correct boundary model, and \cite{dolev2022holography} argued this can be made precise in a lattice regularization of the CFT.\footnote{In more detail, the authors assume a lattice approximation of the holographic CFT exists, and that this lattice approximation captures bulk physics up to a radial cut-off set by the lattice spacing. One interesting recent attempt to construct such lattice descriptions is given in \cite{kohler2019toy,apel2022holographic}.} 
Applied in AdS$_{2+1}$, our construction reaches the same conclusion without introducing a lattice. 
Both constructions can be used to relate bulk computation to non-local computation exploiting finite dimensional entangled states: \cite{dolev2022holography} inherits this from the lattice description, while our approach uses the relation between smooth max entropy and von Neumann entropy in holographic states \cite{bao2019beyond} to justify a finite dimensional approximation. 

By directly relating bulk interactions and non-local computation, these constructions strengthen a tension raised earlier \cite{may2020holographic}.
Assuming a computer can be built in the bulk of AdS, and using that computations happening in the bulk can be reproduced as non-local computations using entanglement related to the area of the bulk region supporting the computation, we expect a large class of computations can be performed efficiently (using polynomial entanglement) in the non-local form. 
In contrast, in the study of non-local computation, it is only known how to implement very low complexity operations using polynomial entanglement. 
In fact, the security of a proposed cryptographic scheme known as quantum position-verification \cite{chandran2009position,kent2011quantum,buhrman2014position} relies on there being low complexity operations requiring large entanglement. 
By removing the need to consider augmented non-local computations in the AdS/CFT context, our construction sharpens this tension between the security of QPV and the ability to build a computer in the bulk of AdS. 

\vspace{0.2cm}
\noindent \textbf{Outline of our article}
\vspace{0.2cm}

In section \ref{sec:ads-bcft} we describe the AdS/BCFT correspondence, our prescription for relating entangled BCFT states to bulk AdS geometries. 
In section \ref{sec:tasks} we review an earlier framework for relating quantum information processing tasks in the bulk and boundary perspectives, into which we can fit our construction.

In section \ref{sec:BTZholes} we begin discussing behind the horizon interactions. 
Section \ref{sec:generaldblackholecase} gives our general construction relating behind the horizon interactions and non-local computation in the context of black holes. 
Section \ref{sec:BTZsolutionPeriodic} applies the construction to the BTZ black hole, where we can solve explicitly for the relevant extremal surfaces. 
Section \ref{sec:BTZsolutionWithBranes} repeats this discussion for the planar black hole ended by ETW branes. 

In section \ref{sec:signature} we make use of our construction to point out a boundary signature of `early' interior collisions. 
We find that scattering in the interior before an extremal surface leads to a large mutual information between associated boundary regions. 

In section \ref{sec:finitedmodel} we point out that the entanglement between two BCFTs can be well approximated by a state in a finite dimensional Hilbert space. 
Using our BCFT based construction, this allows a stronger connection between existing bounds on non-local computation given in the quantum information literature, typically proven for finite dimensional systems, and constraints on interactions happening in a holographic geometry. 

In section \ref{sec:AdS2+1computations} we revisit the global AdS$_{2+1}$ setting, discussed in earlier works \cite{may2019quantum,may2020holographic,may2021quantum,dolev2022holography}.
Here, we point out that ETW brane techniques allow a new perspective on the relationship between bulk interactions and non-local computation. 
The principle claim is that interactions happening in the bulk can be reproduced as non-local computations, using entanglement related to the size of the region in which they occur. 
For large regions, the entanglement needed approaches the area of the region, up to an additive constant. 
These points were discussed earlier but we remove the heuristics of the previous argument. 
In \ref{sec:globalhigherdim} we briefly discuss how to extend the connection between non-local computation and bulk interaction to global AdS$_{d+1}$ for $d>2$. 

In section \ref{sec:discussion} we conclude with some remarks and open questions. 

\vspace{0.2cm}
\noindent \textbf{Summary of notation:}
\vspace{0.2cm}

\noindent Spacetime notation:
\begin{itemize}
    \item We use italic capital letters $\mathcal{A}, \mathcal{B}, \mathcal{C}, \dots$ for boundary spacetime regions.
    \item The entanglement wedge of a boundary region $\mathcal{A}$ is denoted by $E_{\mathcal{A}}$. 
    \item The quantum extremal surface associated to region $\mathcal{A}$ is denoted $\gamma_{\mathcal{A}}$.
    \item We use plain capital letters $A, B, C, \dots$ to refer to bulk spacetime regions.
    \item We use $J^\pm(A)$ to denote the causal future or past of region $A$ taken in the bulk geometry, and $\hat{J}^\pm(\mathcal{A})$ to denote the causal future or past of $\mathcal{A}$ within the boundary spacetime.
\end{itemize}
Quantum notation:
\begin{itemize}
    \item We use capital letters to denote quantum systems $A,B,C,...$
    \item We use boldface, script capital letters for quantum channels, $\mathbfcal{N}(\cdot)$, $\mathbfcal{T}(\cdot)$,...
    \item We use boldface capital letters to denote unitaries or isometries, $\mathbf{U}, \mathbf{V},...$
\end{itemize}

\section{Preliminaries}\label{sec:prelims}

\subsection{AdS/BCFT primer}\label{sec:ads-bcft}

One of our constructions exploits the AdS/BCFT correspondence \cite{takayanagi2011holographic,fujita2011aspects}, which relates holographic CFTs defined on manifolds with a boundary to AdS geometries ended by end-of-the-world branes. 
One way to understand why these solutions appear in our constructions is that we will consider the state on subregions of a CFT, then want some controlled way to consider a purification of that subregion with a well defined bulk geometry. 
Introducing CFT boundary conditions at the edge of the subregion we are considering is one way to do this. 
In this brief section we introduce the needed elements of AdS/BCFT. 

A BCFT is a conformal field theory living on a manifold with boundary, along with a conformally invariant boundary condition. 
For appropriate BCFTs, the AdS/BCFT correspondence conjectures a bulk dual description, which consists of an asymptotically AdS region along with an extension of the CFT boundary into the bulk as an ETW brane. 
To avoid confusion with the bulk-boundary language of the AdS/CFT correspondence, we will refer to the CFT boundary as the \emph{edge}. 
The bulk spacetime and brane are described by an action
\begin{align}\label{eq:ETWaction}
    I_{\text{bulk}}+I_{\text{brane}} & = \frac{1}{16\pi G_N} \int d^{d+1}x \,\sqrt{g} (R-2\Lambda+L_{\text{matter}}) \notag \\ & \qquad + \frac{1}{8\pi G_N} \int_\mathcal{B} d^dy\,\sqrt{h} (K + L_{\text{matter}}^\mathcal{B})\;, 
\end{align}
where $L_{\text{matter}}$ and $L_{\text{matter}}^\mathcal{B}$ are matter Lagrangians for fields in the bulk and brane respectively. 
$R$ is the Ricci curvature and $\Lambda$ the bulk cosmological constant, while $K$ is the trace of the extrinsic curvature of the brane,
\begin{equation}
    K_{ab} = \nabla_a n_b\;,
\end{equation}
for outward normal $n_j$ to $\mathcal{B}$, and $a, b$ refer to brane coordinates $y^a$.
This action leads to Einstein's equations in the bulk, along with the boundary condition
\begin{align}\label{eq:branegeneralBC}
     -\frac{1}{8\pi G_N} (K_{ab}-Kh_{ab}) = T_{ab}^\mathcal{B}\;.
\end{align}
This AdS/BCFT model should be understood as a bottom-up model of concrete holographic dualities.
For example one can set conformally invariant boundary conditions on $\mathcal{N}=4$ SYM and study a holographic dual \cite{aharony2011near,assel2011holographic,bachas2018massive,van2021finding}. 
In these models the bulk geometry has compact dimensions that degenerate somewhere, which corresponds in the bottom-up model to the placement of the ETW brane.
We will be most interested in on-shell action and entropy calculations, where the bottom-up model reproduces universal CFT results \cite{takayanagi2011holographic}. 
More refined probes of these AdS/BCFT models may be more problematic \cite{reeves2021looking}, though our arguments don't rely on the models being correct for those observables.\footnote{In more detail, \cite{reeves2021looking} studies Lorentzian BCFT correlation functions that probe the (putative) brane geometry in the bottom-up model given here.
They find the singularity structure of these correlators doesn't match that expected from bulk causality unless an apparently unnatural condition is imposed on the BCFT spectrum.
In contrast, our setting involves a bulk process that remains far away from the branes, and we only rely on the bottom up model to determine the placement of minimal surfaces and to understand when a particular geometry is the relevant bulk saddle.}  

The Ryu-Takayanagi formula gives a method for computing boundary entropy in terms of bulk degrees of freedom. 
In one of its modern forms, this can be expressed as
\cite{engelhardt2015quantum}
\begin{align}\label{eq:QESformula}
    S(A) =  \min_{\gamma_{ext}} \text{ext}_{\gamma\in \text{Hom}(A) }\left(\frac{\text{area}(\gamma)}{4G_N} + S_{bulk}(E_\gamma) \right).
\end{align}
The set $\text{Hom}(A)$ is the set of codimension $2$ spacelike surfaces such that there exists a codimension 1 spacelike surface $E_\gamma$ satisfying
\begin{align}
    \partial E_\gamma = \gamma \cup A.
\end{align}
We say that $\gamma\in \text{Hom}(A)$ are \emph{homologous} to $A$. 
The quantity $S_{bulk}(E_\gamma)$ is the von Neumann entropy of the bulk subregion $E_{\gamma}$. 
In AdS/BCFT, the Ryu-Takayanagi formula continues to calculate the entropy of boundary subregions, provided the homology condition is appropriately adapted \cite{sully2021bcft}.
The appropriate definition of homologous in the presence of ETW branes is that there needs to exist a spacelike codimension 1 surface $E_\gamma$ such that 
\begin{align}\label{eq:S}
    \partial E_\gamma = \gamma \cup A \cup b
\end{align}
where $b$ is allowed to be any portion of the ETW brane. 

Given a subregion of the boundary $A$, it is natural to ask if a subregion of the bulk is recorded into $A$. 
To make this question more precise, we should introduce a choice of bulk subspace, which we refer to as the code-space and label $\mathcal{H}_{code}$. 
The subspace $\mathcal{H}_{code}$ might for instance be specified by a particular choice of bulk geometry, along with some qubits distributed spatially across the bulk.
Then, assume we are told the bulk degrees of freedom are in a state within $\mathcal{H}_{code}$, and we are given the degrees of freedom on subregion $A$. 
What portion of the bulk degree's of freedom can we recover?

Answering this question is related closely to the RT formula, in both AdS/CFT and AdS/BCFT. 
In particular, the portion of the bulk we can recover if we know the bulk state in $\mathcal{H}_{code}$ is given by \cite{hayden2019learning,akers2019large}
\begin{align}
    E_A \equiv \bigcap_{\psi\in \mathcal{H}_{code}} E_{\gamma_A}.
\end{align}
That is, for each state in the code space we find where the RT surface $\gamma_A$ sits, and define the corresponding bulk subregion $E_{\gamma_A}$. 
Then, we define the intersection of all such surfaces, considering all states in the code-subspace. 
Note that in this procedure we should include mixed states of the code-space. 
The resulting region is the portion of the bulk degrees of freedom we can recover, if we know nothing about which state in the code-space the full bulk is in. 
This region is sometimes referred to as the \emph{reconstruction wedge} of region $A$, defined relative to the code-space $\mathcal{H}_{code}$.

Given that it is possible to recover information inside the reconstruction wedge, we can also ask what explicit operation recovers the code space from the CFT degrees of freedom.
Given a global map from the bulk subspace $\mathcal{H}_{code}$ to the boundary Hilbert space, it was understood in \cite{cotler2019entanglement} how to construct such a recovery channel.
Note that in this construction, a single choice of recovery channel works correctly for the entire code-space. 

\begin{figure}
    \centering
    \begin{tikzpicture}
    
    \draw[ultra thick] (-4,0) -- (0,0);
    \draw[ultra thick,cyan] (0,0) -- (2.5,2.5);
    \draw[thick,blue,domain=90:45] plot ({2*cos(\x)},{2*sin(\x)}); 
    \draw[dashed] (0,0) -- (0,4);
    \node at (0.9,2.2) {\Large{$\Theta$}};
    
    \end{tikzpicture}
    \caption{Geometry dual to the vacuum state of a BCFT on a half line.}
    \label{fig:angle}
\end{figure}
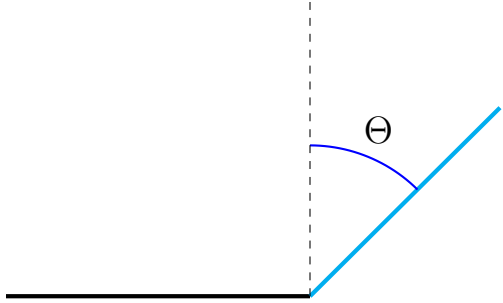

To illustrate the AdS/BCFT correspondence, we show a simple AdS/BCFT solution in figure \ref{fig:angle}. 
There, the BCFT is defined on $\mathbb{R}^-\times \mathbb{R}$, though we supress the timelike $\mathbb{R}$ direction in the figure.  
We have taken the bulk matter Lagrangian to be zero, and the brane matter Lagrangian to be
\begin{align}
    L^{\mathcal{B}}_{matter} = -T.
\end{align}
The parameter $T$ is referred to as the tension. 
The solution can be described by
\begin{align}
    ds^2 = \frac{\ell^2}{z^2}(dz^2 - dt^2 + dx^2)
\end{align}
with an ETW brane along an $AdS_{1+1}$ slice $\frac{x}{z} = \tan \Theta$,
where the angle $\Theta$ is related to the tension parameter as
\begin{align}
    \sin \Theta = T \ell .
\end{align}
Notice that consistent solutions require
\begin{align}
    -\frac{1}{\ell} \leq T\leq \frac{1}{\ell}.
\end{align}
In the CFT, the tension parameter is related to the boundary entropy, a measure of the number of degrees of freedom localized to the edge. 
The boundary entropy is expressed as $S_b=\log g$, and is related to the tension by
\begin{align}
    \log g_B = \frac{\ell}{4G_N}\text{arctanh}(T\ell).
\end{align}
Qualitatively, increasing the boundary entropy adds degrees of freedom to the CFT, and corresponding increases the angle $\Theta$, adding a portion of bulk spacetime. 

\subsection{Holographic quantum tasks with input and output regions}\label{sec:tasks}

In this section we review an operational perspective on AdS/CFT introduced in \cite{may2019quantum} and subsequently developed in \cite{may2020holographic,may2021quantum, may2022connected}. 
The basic idea is to consider quantum information processes that can be interpreted in the bulk and boundary perspectives, and to reason about how the two descriptions of these processes inform each other. 
An up to date and more complete description of this perspective can be found in \cite{may2022connected}. 
Here we briefly review a few ingredients which contextualize our discussion. 

We will consider information processing tasks that have inputs and outputs distributed in spacetime.
To describe this task succinctly, it is helpful to introduce two agencies, Alice and Bob, who interact to carry out the task.\footnote{Note that introducing Alice and Bob is only a linguistic convenience: we could rephrase our settings in the language without any notion of observer or agent.} 
We will label the input locations by $\mathcal{C}_i$ where Bob gives Alice quantum systems $A_i$. 
These may be entangled with some reference $R$, which we take to be held by Bob. 
Alice carries out the task by manipulating the input systems and preparing a set of output systems $B_i$, which she returns to Bob at another set of spacetime locations $\mathcal{R}_i$. 
Bob then makes a measurement on the $B=B_1B_2...B_n$ system with elements $\{\Lambda_B,\mathcal{I}_B-\Lambda_B \}$ and we declare Alice successful if she obtains outcome $\Lambda_B$. 
Following Kent \cite{kent2012quantum}, who initially formalized this notion, we refer to these scenarios as \emph{relativistic quantum tasks}. 

The input and output locations will be extended regions in spacetime, though we sometimes idealize these as spacetime points if we wish to take the regions to be small. 
We will consider relativistic quantum tasks in the bulk and boundary perspectives in AdS/CFT. 
By comparing the two perspectives, we can gain interesting insights into the AdS/CFT correspondence and into quantum information processing.

The key ingredient in relating bulk and boundary perspectives on relativistic quantum tasks is the notion of the reconstruction wedge, which we reviewed briefly in the last section. 
In particular, given a task defined in the bulk geometry with input or output region ${X}_i$, this will correspond to a task in the boundary which has an input or output region containing ${X}_i$ in its reconstruction wedge. 
To make this more precise, consider a task $\mathbf{T}$ defined in the bulk with input regions $\{{C}_i\}_i$ and output regions $\{{R}_i\}_i$. 
In the boundary, consider a task with input regions $\{\mathcal{C}_i\}_i$, $\{\mathcal{R}_i\}_i$ such that 
\begin{align}
    C_i &\subseteq E_{\mathcal{C}_i}, \nonumber \\
    R_i &\subseteq E_{\mathcal{R}_i}. \nonumber
\end{align}
Note that in general there can be many such choices of boundary regions, we can take any one of them. 
The boundary task with the input and output regions defined in this way is taken to have the same input and output systems $A_i$, $B_i$, and the same choice of measurement that defines the success or failure of the task. 
We call the resulting boundary task $\hat{\mathbf{T}}$.

The tasks $\mathbf{T}$ and $\hat{\mathbf{T}}$ are related in a simple way.
In particular, we can notice that a strategy for completing the task in the bulk that succeeds with probability $p_{suc}(\mathbf{T})$ implies the existence of a corresponding strategy in the boundary that succeeds with the same probability. 
This is because given access to boundary regions $\mathcal{C}_i$, $\mathcal{R}_i$, Alice can encode her inputs into bulk regions $C_i$, allow boundary time evolution (which in the bulk picture implements the strategy for completing the task), then recover the outputs $B_i$ from output regions $\mathcal{R}_i$. 
We can summarize this as
\begin{align}
    p_{suc}(\mathbf{T}) \leq p_{suc}(\hat{\mathbf{T}}).
\end{align}
This is the starting point for the various implications for AdS/CFT that can extracted from studying relativistic quantum tasks \cite{may2020holographic,may2021bulk,may2021holographic,may2021quantum,may2022connected}. 
This can also be strengthened in an interesting way, as pointed out in \cite{may2022connected}, although we won't use the stronger version here.  

\section{Computation inside black holes}\label{sec:BTZholes}

In this section, we show that certain computations happening inside of black holes can be reproduced as non-local quantum computations.
More precisely, computations inside of a particular subregion of the black hole, defined below, can be reproduced in the non-local form of figure \ref{fig:non-localcomputation} using a resource system with mutual information given by the black hole area. 
For concreteness we discuss planar black holes, but a similar construction works for bulk duals of CFTs with compact spatial directions (we discuss the BTZ black hole case in more detail below). 
In some special settings, the region where interactions can be reproduced as non-local computations reaches all the way to the future boundary of the spacetime, but this is not always the case. 

\subsection{General construction}\label{sec:generaldblackholecase}

\begin{figure}
    \centering
    \subfloat[\label{fig:sidewayscylinder}]{
    \tdplotsetmaincoords{15}{0}
    \begin{tikzpicture}[scale=0.75,tdplot_main_coords]
    \tdplotsetrotatedcoords{0}{-30}{0}
    
    \begin{scope}[tdplot_rotated_coords]

    \draw (0,{2*cos(190)},{2*sin(190)}) -- (5,{2*cos(190)},{2*sin(190)});
    \draw (0,{2*cos(10)},{2*sin(10)}) -- (5,{2*cos(10)},{2*sin(10)});

    \draw[opacity=0.75] (0,0,2) -- (5,0,2);

    \begin{scope}[canvas is yz plane at x=5]
    \draw[thick,black,opacity=0.5] (0,0) circle[radius=2] ;
    \end{scope}

    \begin{scope}[canvas is yz plane at x=0]
    \draw[thick,black] (0,0) circle[radius=2] ;
    \end{scope}

    \begin{scope}[canvas is yz plane at x=5]
    \draw [domain=190:370,black] plot ({2*cos(\x)}, {2*sin(\x)});
    \end{scope}
    
    \draw (0,0,-2) -- (5,0,-2);

    \draw[->] (0,2.5) -- (1,2.5);
    \node[above] at (0,2.5) {$X_1$};

    \begin{scope}[canvas is yz plane at x=5]
    \draw [domain=270:290,black,->] plot ({2.4*cos(\x)}, {2.4*sin(\x)});
    \end{scope}

    \node[right] at (5,0,-2.4) {$\phi$};
    
    \end{scope}
    \end{tikzpicture}}
    \hfill
    \subfloat[\label{fig:box}]{
    \centering
    \tdplotsetmaincoords{15}{0}
    \centering
    \begin{tikzpicture}[scale=0.5,tdplot_main_coords]
    \tdplotsetrotatedcoords{0}{30}{0}
    
    \begin{scope}[tdplot_rotated_coords]

    \draw[thick,black] (0,0,0) -- (0,0,10) -- (0,5,10) -- (0,5,0) -- (0,0,0);
    \draw[thick,black] (-5,0,0) -- (-5,0,10) -- (-5,5,10) -- (-5,5,0) -- (-5,0,0);
    \draw[thick] (-5,0,0) -- (0,0,0) -- (0,5,0) -- (-5,5,0) -- (-5,0,0);
    \draw[thick] (-5,0,10) -- (0,0,10) -- (0,5,10) -- (-5,5,10) -- (-5,0,10);

    \draw[dashed,thick] (-5,0,0) -- (0,5,0);
    \draw[dashed,thick] (-5,5,0) -- (0,0,0);

    \draw[dashed,thick] (-5,0,10) -- (0,5,10);
    \draw[dashed,thick] (-5,5,10) -- (0,0,10);

    \draw[->] (1,0,5) -- (1,0,7);
    \node[right] at (1,0,7) {${X}_1$};

    \draw[->] (1,0,10) -- (1,1.5,10);
    \node[above] at (1,1.5,10) {${t}$};
    
    \end{scope}
    \end{tikzpicture}
    }
    \caption{The planar black hole. (a) A path integral on the cylinder $S^1\times \mathbb{R}^d$. (b) Wick rotating the angular coordinate $\phi\rightarrow it$ of the Euclidean geometry on the left we obtain the Lorentzian planar black hole.}
    \label{fig:boxgeometry}
\end{figure}
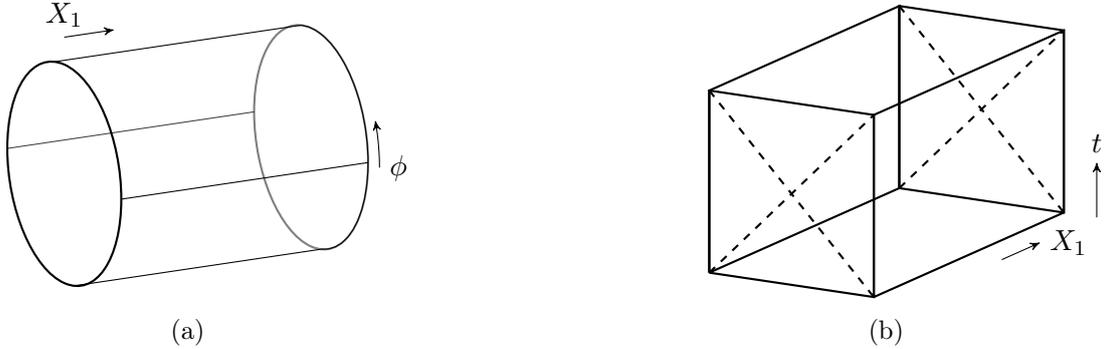

The metric for the AdS$_{d+1}$ planar black brane can be written as \cite{hartman2013time}
\begin{align}\label{eq:generaldBHmetric}
    ds^2 = -(\ell g(\rho))^2 dt^2 + \ell^2 d\rho^2 +(\ell h(\rho))^2\sum_{i=1}^{d-1} dX^2_{i}
\end{align}
where
\begin{align}
    h(\rho) &= \frac{2}{d} \left( \cosh\left( \frac{d\rho}{2}\right)\right)^{2/d} \nonumber \\
    g(\rho) &= h(\rho) \tanh\left( \frac{d\rho}{2}\right).
\end{align}
These coordinates cover one exterior region for $\rho\geq 0$. 
The temperature of the black hole has been fixed at $1/2\pi$. 
A Penrose diagram showing the maximally extended geometry is shown in figure \ref{fig:boxgeometry}. 
The boundary dual is understood to be copies of a holographic CFT, placed in the thermofield double state. 
Each CFT lives on a $d$ dimensional Minkowski space. 

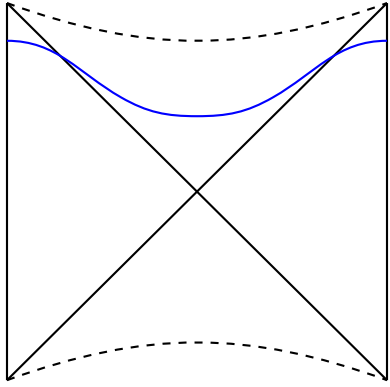
\begin{figure}
    \centering
    \begin{tikzpicture}

    \draw[thick] (0,0) -- (5,5);
    \draw[thick] (5,0) -- (0,5);
    \draw[thick] (0,0) -- (0,5);
    \draw[thick] (5,0) -- (5,5);

    \draw[thick,dashed] (0,0) to [out=20,in=160] (5,0);
    \draw[thick,dashed] (0,5) to [out=-20,in=-160] (5,5);

    \draw[blue,thick] (0,4.5) to [out=0,in=150] (1.5,3.75) to [out=-30,in=-180] (2.5,3.5) to [out=0,in=-150] (3.5,3.75) to [out=30,in=180] (5,4.5);
    
    \end{tikzpicture}
    \caption{Penrose diagram of the planar black brane. $d-1$ planar spatial directions are supressed. An extremal surface (blue) anchored to the boundary at $x_0=0$ threads through the black hole. For a $2+1$ dimensional bulk the future boundary and blue surface are flat in the Penrose diagram.}
    \label{fig:planarPenrose}
\end{figure}

We will consider the entanglement wedges of half spaces of the combined CFTs, defined by $X_1 < 0$ and $X_1 > 0$. 
As discussed in \cite{hartman2013time}, the corresponding minimal surfaces thread through the black hole. 
We illustrate this in figure \ref{fig:planarPenrose}.
The explicit form of this surface is discussed there. 
We will give explicit formula for this surface in the low dimensional case treated below, but for now only need to note that this extremal surface exists.
We will call this the connected extremal surface. 

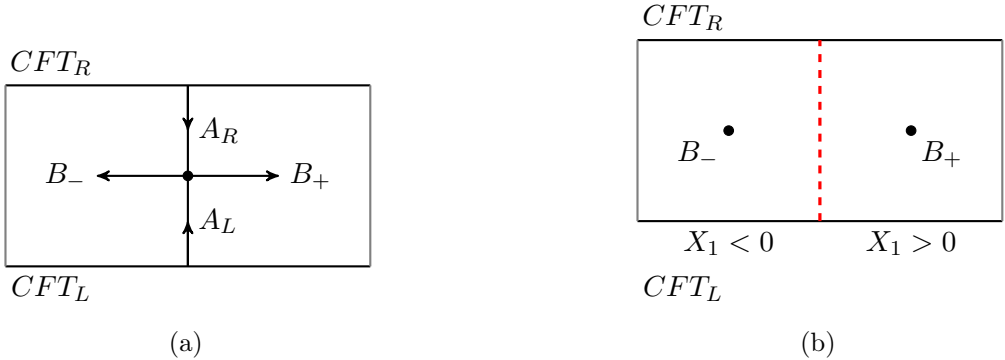
\begin{figure*}
    \centering
    \begin{subfigure}{0.45\textwidth}
    \centering
    \begin{tikzpicture}[scale=0.6]

    \draw[thick,black] (0,0) -- (8,0);
    \draw[thick,black] (0,4) -- (8,4);
    \draw[thick,gray] (0,0) -- (0,4);
    \draw[thick,gray] (8,0) -- (8,4);
    \draw[thick,black] (4,0) -- (4,4);

    \draw[thick,->] (4,0) -- (4,1) node[anchor=west]{$A_L$};
    \draw[thick,->] (4,4) -- (4,3) node[anchor=west]{$A_R$};

    \draw[thick,->] (4,2) -- (2,2)node[anchor=east]{$B_-$};
    \draw[thick,->] (4,2) -- (6,2)node[anchor=west]{$B_+$};

    \draw[black] plot [mark=*, mark size=3] coordinates{(4,2)};

    \node[below] at (1,0) {$CFT_L$};
    \node[above] at (1,4) {$CFT_R$};

    \end{tikzpicture}
    \caption{}
    \label{fig:infallingsystemsBH}
    \end{subfigure}
    \hfill
    \begin{subfigure}{0.45\textwidth}
    \centering
    \begin{tikzpicture}[scale=0.6]
 
    \draw[thick,black] (0,0) -- (8,0);
    \draw[thick,black] (0,4) -- (8,4);
    \draw[thick,gray] (0,0) -- (0,4);
    \draw[thick,gray] (8,0) -- (8,4);
    \draw[very thick,red, dashed] (4,0) -- (4,4);

    \draw[black] plot [mark=*, mark size=3] coordinates{(2,2)} node[anchor=north east]{$B_-$};
    \draw[black] plot [mark=*, mark size=3] coordinates{(6,2)} node[anchor=north west]{$B_+$};
    
    \node[below] at (1,-1) {$CFT_L$};
    \node[above] at (1,4) {$CFT_R$};
    \node[below] at (2,0) {$X_1<0$};
    \node[below] at (6,0) {$X_1>0$};
    \end{tikzpicture}
    \caption{}
    \label{fig:outgoingsystemsBH}
    \end{subfigure}
    \caption{Illustration of our thought experiment reproducing bulk interactions as non-local computations. Alice$_L$ and Alice$_R$ hold the left and right CFTs of the solution shown in figure \ref{fig:boxgeometry}. (a) The Alice's receive the inputs to the computation, and throw their respective systems into the black hole. The systems interact inside of the black hole interior. (b) After the two CFTs have evolved to time $T$, Alice$_L$ and Alice$_R$ use a single, simultaneous round of communication to redistribute systems so that Alice$_L$ holds the $X^1>0$ portion of both CFTs, and Alice$_R$ holds both of the $X^1<0$ intervals. The portion of the bulk that can be recovered from the two $X_1>0$ intervals includes $B_+$, and the systems recoverable from the two $X^1<0$ intervals includes $B_-$. From the boundary perspective this process is described as a circuit with the form shown in figure \ref{fig:non-localcomputation}.}
    \label{fig:BTZprotocol}
\end{figure*}

The entanglement between the two CFTs corresponding to our planar black hole is infinite. 
Later it will be important to render this entanglement finite, and we discuss two methods for doing so. 
First, we can periodically identify the $X_1$ direction over $[-X^1_0,X_0^1]$. 
Alternatively, we can introduce ETW branes at $X^1=\pm X^1_0$. 
We discuss both strategies, and complications that arise there, in the sections below. 
In this section we continue with the idealized planar case. 

To begin our construction, consider quantum systems $A_L$, $A_R$ which are initially stored within the left and right asymptotic regions of the planar black hole, respectively. 
These systems fall into the black hole, interact, and produce output systems $B_+$, $B_-$. 
We then have these systems travel in the $\hat{X}_1$ and $-\hat{X}_1$ directions, respectively. 
The overall process is described in figure \ref{fig:BTZprotocol}. 
We take all relevant systems to consist of fewer qubits than the black hole entropy.

To describe this process in the boundary perspective, we first note that we can take as input regions $\mathcal{C}_L$, $\mathcal{C}_R$ the entire left and right CFTs. 
These then contain the inputs $A_L$ and $A_R$ in their reconstruction wedges, which will be the left and right exterior regions. 
For output regions, we choose a time $T$ in the CFT, and consider the CFT regions
\begin{align}\label{eq:outputregions}
    \mathcal{R}_- &= \{ p: t=T, X_1<0\}, \nonumber \\
    \mathcal{R}_+ &= \{ p: t=T, X_1>0\} .
\end{align}
In words $\mathcal{R}_-$ is the $X_1<0$ portion of both CFTs taken at time $T$, $\mathcal{R}_+$ is the $X_1>0$ portions of both CFTs taken at the same time. 

The boundary of the reconstruction wedges of $\mathcal{R}_+$ and $\mathcal{R}_-$ is defined by the extremal surface anchored to the two boundary cuts $\gamma_+$, $\gamma_-$, defined at $X_1=0, t=T$ in the two CFTs.
In the planar black hole this threads through the black hole interior as shown in figure \ref{fig:planarPenrose}.
Consider an interaction that takes place in the bulk subregion
\begin{align}
    J_T = J^-(E_{\mathcal{R}_+}) \cap J^-(E_{\mathcal{R}_-}) \cap J^+(E_{\mathcal{C}_L}) \cap J^+(E_{\mathcal{C}_R})
\end{align}
which we can also express as
\begin{align}
    J_T = J^-(E_{\mathcal{R}_+}) \cap J^-(E_{\mathcal{R}_-}) \cap I
\end{align}
where $I$ denotes the black hole interior. 
We refer to $J_T$ as the \emph{scattering region}. 
If the interaction occurs inside this region, then the output systems $B_-$ and $B_+$ will each enter one of the wedges $E_{\mathcal{R}_-}$ and $E_{\mathcal{R}_+}$. 

Interactions happening inside the scattering region can be reproduced as non-local computations. 
To see this, consider the non-local computation circuit shown in figure \ref{fig:non-localcomputation}. 
Relating that picture to our setting, the left and right ends of the shared entangled state represent the left and right CFTs. 
The local operations on the left and right at the first time step can be used to insert the inputs $A_L$ and $A_R$ into the bulk, and allow the left and right CFTs to evolve under time evolution. 
Then, we consider recovery operations that act on $\mathcal{R}_\pm$. 
Since systems $B_\pm$ have been brought into the corresponding reconstruction wedges $E_{\mathcal{R}_\pm}$, these systems can be extracted from the bulk by acting on $\mathcal{R}_\pm$. 
The swap operation at the middle level then collects both the $+$ halves and both the $-$ halves of the CFTs, so that the final round operations can be taken to be the necessary recovery operations. 
Thus we can reproduce the bulk computation in the form of a non-local computation. 
Further, the needed entangled state was exactly the thermofield double, which has a black hole's entropy worth of mutual information. 

We can also express the relationship to non-local computation in more operational language. 
We consider two agents, Alice$_L$ and Alice$_R$, who initially hold the left and right CFTs respectively. 
Alice$_L$ acts on her CFT to insert $A_L$ into the bulk, Alice$_R$ acts on her CFT to insert $A_R$ into the bulk. 
Both locally apply the time evolution operators for their CFTs. 
Next, Alice$_L$ and Alice$_R$ divide their CFT degrees of freedom and exchange a round of communication to bring the $\mathcal{R}_+$ degrees of freedom to Alice$_R$ and $\mathcal{R}_-$ degrees of freedom to Alice$_L$. 
They then each apply the appropriate recovery operations to produce the $B_-$ and $B_+$ systems. 

Beyond the circuit \ref{fig:non-localcomputation} having the form of a non-local computation, our construction also shows it has an additional property. 
In particular, a non-local circuit capturing a bulk interaction can always be taken to have fixed recovery channels as its second round operations. 
This means bulk computation must be reproducible in a stricter form than the usual form of a non-local computation. 
This property also appears in the construction in \cite{dolev2022holography}, though isn't noted explicitly. 
We comment on this property further in the discussion.

\subsection{Example: BTZ}\label{sec:BTZsolutionPeriodic}

In this section and the next we give two constructions, building on the above, which relate bulk interactions inside of black holes to non-local computation circuits, now with finite entanglement. 
We give these finite entanglement constructions in bulk AdS$_{2+1}$ but similar constructions work in higher dimensions. 

We first consider the spherical BTZ black hole. 
This is related to the planar black hole construction given in the last section by making a periodic identification so that $X^1\in [-\pi,\pi]$. 
In doing so, the Euclidean path integral shown in figure \ref{fig:sidewayscylinder} is now on a torus $[0,\beta)\times [-\pi,\pi]$. 
The bulk saddle for this torus boundary is known as the $\text{SL}(2,\mathbb{Z})$ family of black holes \cite{maloney2010quantum, maldacena1999ads3}. 
The saddle we need for our protocol is the Euclidean BTZ black hole \cite{carlip1995aspects}, which is dominant when $\beta < 2\pi$. 
We assume this for the remainder of this section. 

After Wick rotation, we arrive at the standard BTZ black hole geometry, which is dual to a thermofield double state with temperature $1/\beta$ defined on two periodic CFTs $[0, 2\pi)$ on the boundary. 
The BTZ black hole metric in Kruskal coordinates \cite{shenker2014black} is
\begin{align}\label{eq:kruskal_btz}
    ds^2 = \frac{-4\ell^2 dudv+R^2(1-uv)^2d\phi^2}{(1+uv)^2}
\end{align}
with the convention that the right exterior has $u<0$ and $v>0$, the boundaries are at $uv=-1$, and the singularities are at $uv=1$. 
The parameter $R$ is related to $\beta$ by $R=2\pi \ell/\beta$.

This Kruskal coordinates metric can be straightforwardly transformed to,
\begin{align}\label{eq:periodicBTZ}
ds^2 = \frac{\ell^2}{\cos^2(w)}\left(-ds^2+dw^2+\cos^2(s)dX^2\right).
\end{align}
where $X$ is periodic over $(-X^0,X^0)$ with $X^0=2\pi^2/\beta$. 
We choose these coordinates because they both simplify extremal surface calculations and allow us to re-use our calculation for the planar setting, discussed in the next section. 
The coordinate change relating this metric to equation \ref{eq:kruskal_btz} is given in appendix \ref{appendix:periodic_BTZ}.

To relate bulk interaction to non-local computation, consider the extremal surface anchored to the points $X=0,s=T, w=\pm \pi/2$ and $X=X^0,s=T, w=\pm \pi/2$. We require four extremal surface anchor points for our setup since our protocol requires partitioning the boundary into a "left" and "right" half of the (Wick-rotated) torus. There are two candidate extremal surfaces: one is comprised of two pieces that each connect from one boundary to the other and thread through the black hole, while the other scenario is two pieces, each of which connects two anchor points from the same boundary and does not go through the black hole. 
The surface $\gamma_T$ that threads through the black hole is,
\begin{align}\label{eq:BTZsurface}
    w &= \arcsin(\tanh \lambda),\nonumber \\
    s &= T
\end{align}
The connected surface consists of two of these, one at $X=0$ and one at $X=X^0$.
In appendix \ref{appendix:periodic_BTZ}, we find the areas corresponding to both the surface which connects the boundaries, and the surface which does not connect the boundaries. 
The connected surface has area
\begin{align}
    A_c[T] = 4\ell \log \left( \frac{2}{\epsilon}\right),
\end{align}
and the disconnected one has area
\begin{align}
    A_d[T] = 4 \ell \log\left( \frac{2 \sinh\left(\frac{|X^0|}{2}\right) \cos(T)}{\epsilon}\right).
\end{align}
These give that the area difference is
\begin{align}\label{eq:periodic_deltaA}
    \Delta A = 4 \ell \log\left( \sinh\frac{|X^0|}{2} \cos T \right)
\end{align}
and the connected surface is minimal when
\begin{align}\label{eq:periodic_minsurfacecondition}
    \sinh \frac{X^0}{2} \geq \sec T.
\end{align}
If we adapt the general discussion in section \ref{sec:generaldblackholecase} to the periodic case, we see that in order to reproduce bulk interactions as non-local computations, we require both the geometry to be in the black hole saddle, i.e. $\beta < 2\pi$, and the condition above.
Using that $X^0=2\pi^2/\beta$, this gives that
\begin{align}\label{eq:sphericaltempcondition}
    \sinh \frac{\pi^2}{\beta} \geq \sec T.
\end{align}
Notice also that we can only reproduce as non-local computations those interactions that happen in the past of the extremal surface threading through the black hole. 
For any fixed temperature black hole then there is a latest time, given by taking an equality in \ref{eq:sphericaltempcondition}, above which we cannot reproduce bulk interactions as non-local computations. 

\subsection{Example: planar BTZ with ETW branes}\label{sec:BTZsolutionWithBranes}

\begin{figure}
    \centering
    \subfloat[\label{fig:connectedbranes}]{
    \tdplotsetmaincoords{15}{0}
    \begin{tikzpicture}[scale=1,tdplot_main_coords]
    \tdplotsetrotatedcoords{0}{-30}{0}
    
    \begin{scope}[tdplot_rotated_coords]

    \draw (0,{2*cos(190)},{2*sin(190)}) -- (5,{2*cos(190)},{2*sin(190)});
    \draw (0,{2*cos(10)},{2*sin(10)}) -- (5,{2*cos(10)},{2*sin(10)});

    \draw[opacity=0.75] (0,0,2) -- (5,0,2);

    \begin{scope}[canvas is yz plane at x=5]
    \draw[thick,black,opacity=0.5] (0,0) circle[radius=2] ;
    \end{scope}

    \foreach \i in {1,...,100}
    {
    \begin{scope}[canvas is yz plane at x={5*\i/100}]
    \draw[gray,opacity=0.5] (0,0) circle[radius={0.16*(5*\i/100 - 2.5)^2+1}] ;
    \end{scope}
    }

    \begin{scope}[canvas is yz plane at x=0]
    \draw[thick,black] (0,0) circle[radius=2] ;
    \end{scope}

    \begin{scope}[canvas is yz plane at x=5]
    \draw [domain=190:370,black] plot ({2*cos(\x)}, {2*sin(\x)});
    \end{scope}
    
    \draw (0,0,-2) -- (5,0,-2);

    \draw[->] (0,2.5) -- (1,2.5);
    \node[above] at (0,2.5) {$X_1$};

    \begin{scope}[canvas is yz plane at x=5]
    \draw [domain=270:290,black,->] plot ({2.4*cos(\x)}, {2.4*sin(\x)});
    \end{scope}

    \node[right] at (5,0,-2.4) {$\phi$};
    
    \end{scope}
    \end{tikzpicture}}
    \hfill
    \subfloat[\label{fig:disconnectedbranes}]{
    \tdplotsetmaincoords{15}{0}
    \begin{tikzpicture}[scale=1,tdplot_main_coords]
    \tdplotsetrotatedcoords{0}{-30}{0}
    
    \begin{scope}[tdplot_rotated_coords]

    \draw (0,{2*cos(190)},{2*sin(190)}) -- (5,{2*cos(190)},{2*sin(190)});
    \draw (0,{2*cos(10)},{2*sin(10)}) -- (5,{2*cos(10)},{2*sin(10)});

    \draw (0,0,2) -- (5,0,2);
    \draw (0,0,-2) -- (5,0,-2);
    
    \begin{scope}[canvas is yz plane at x=0]
    \draw[thick,black,fill=gray,opacity=0.5] (0,0) circle[radius=2] ;
    \draw[thick,black] (0,0) circle[radius=2] ;
    \end{scope}

    \begin{scope}[canvas is yz plane at x=5]
    \draw[thick,black,fill=gray,opacity=0.5] (0,0) circle[radius=2] ;
    \end{scope}

    \begin{scope}[canvas is yz plane at x=5]
    \draw [domain=190:370,black] plot ({2*cos(\x)}, {2*sin(\x)});
    \end{scope}
    
    \end{scope}
    \end{tikzpicture}
    }
    \caption{Euclidean gravity solutions corresponding to the thermofield double state of two BCFTs on intervals. (a) When the interval is short, the ETW brane connects the two BCFT edges. (b) When the interval is long, the ETW brane is in two pieces, and separately attached to each edge. We Wick rotate the angular $\phi$ coordinate. The solution with two disconnected branes Wick rotates to a planar black hole.}
    \label{fig:branetransition}
\end{figure}
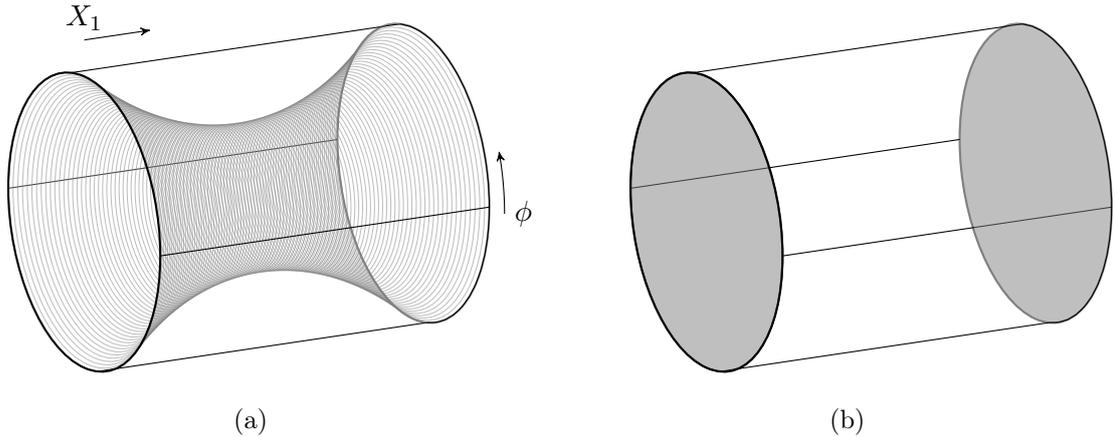

In this section we adapt our construction in section \ref{sec:generaldblackholecase} to a setting where the planar direction is now truncated, so that the geometry is ended by two ETW branes. 
The dual description is a pair of BCFTs living on the spatial interval $[-X^0,X^0]$.
For simplicity, we take the bulk ETW branes to be zero tension. 
We focus on the case of bulk AdS$_{2+1}$ but our construction can be adapted to higher dimensions. 

To construct our planar BTZ black hole solution, we again begin with the Euclidean path integral on a cylinder $[0,2\pi)\times [-X^0,X^0]$ and fix the black hole temperature to be $1/2\pi$.
The bulk solution is global Euclidean AdS, 
\begin{align}
    ds^2 = \ell^2\left(\frac{\rho^2}{\ell^2} +1\right)dX^2 + \frac{d\rho^2}{\frac{\rho^2}{\ell^2}+1} +\rho^2 d\phi^2. 
\end{align}
As studied in \cite{cooper2019black}\footnote{See equation A.24 there. Note that they perform a different Wick rotation, so our black hole occurs for disconnected branes, while the black hole in \cite{cooper2019black} appears for connected branes. The same transition can also be extracted from \cite{fujita2011aspects}.}, at zero tension the ETW branes are in the disconnected configuration of figure \ref{fig:disconnectedbranes} when $X^0\geq \pi/2$, and otherwise the two branes connect.
Explicitly, the branes sit at $X=\pm X^0$.
From here forward, we take $X^0 \geq \frac{\pi}{2}$.

Considering the metric above, Wick rotate $\phi\rightarrow i t$. 
The result is the thermofield double state with temperature $1/2\pi$, defined on two BCFTs on intervals $[-X^0,X^0]$. 
The resulting metric
\begin{align}\label{eq:exteriorBTZ}
    ds^2 = \ell^2\left(\frac{\rho^2}{\ell^2} +1\right)dX^2 + \frac{d\rho^2}{\frac{\rho^2}{\ell^2}+1} - \rho^2 d\phi^2 
\end{align}
covers the exterior region $\rho>0$. 
We can extend this geometry to a full Lorentzian black hole geometry, obtaining 
\begin{align}\label{eq:planarBTZ}
    ds^2 = \frac{\ell^2}{\cos^2(w)}\left(-ds^2 + dw^2 + \cos^2(s) dX^2 \right).
\end{align}
Notice this metric takes the same form as \ref{eq:periodicBTZ}, but now $X$ is finite since the two end of the world branes are located at $X=\pm X^0$. 
The asymptotic regions sit at $w=\pm \pi/2$, and the black hole horizons at $s=\pm w$. 
The coordinate change relating this metric to \ref{eq:exteriorBTZ} is given in appendix \ref{sec:coordinatechanges}.

To relate interaction in our planar bulk to non-local computation, we consider the extremal surface anchored onto $X=0,s=T, w=\pm \pi/2$. There are two candidate extremal surfaces, which are analogous to the ones in the periodic case: a connected surface that threads through the black hole, and a disconnected surface with two pieces that connects each of the boundary anchor points to the ETW brane.
The form of the connected surface $\gamma_T$ is again
\begin{align}
    w &= \arcsin(\tanh \lambda),\nonumber \\
    s &= T.
\end{align}
We find the areas of both this and the brane-anchored surfaces in appendix \ref{appendix:BTZ}, obtaining
\begin{align}
    A_c[T] = 2\ell \log \left( \frac{2}{\epsilon}\right),
\end{align}
and
\begin{align}
    A_d[T] = 2 \ell \log\left( \frac{2 \sinh|X^0| \cos(T)}{\epsilon}\right),
\end{align}
which gives an area difference of
\begin{align}\label{eq:deltaA}
    \Delta A = 2 \ell \log\left( \sinh |X^0| \cos T \right).
\end{align}
We conclude that the connected surface is minimal when
\begin{align}\label{eq:minsurfacecondition}
    \sinh X^0 \geq \sec T.
\end{align}
Adapting the construction in section \ref{sec:generaldblackholecase}, we find that we need both equation \ref{eq:minsurfacecondition} and $X^0\geq \pi/2$ for bulk interactions in the planar BTZ black hole to be reproduced as non-local computations. 
Further, it is interactions that happen in the past of this extremal surface and interior of the black hole can that be reproduced in this way. 
A comment is that if we wish to take $T\rightarrow \pi/2$ so that interactions close to the future boundary of the spacetime can be reproduced, we then require $X^0\rightarrow \infty$, so that the entanglement between the left and right CFTs becomes infinite. 

\section{Boundary signature for early interior collisions}\label{sec:signature}

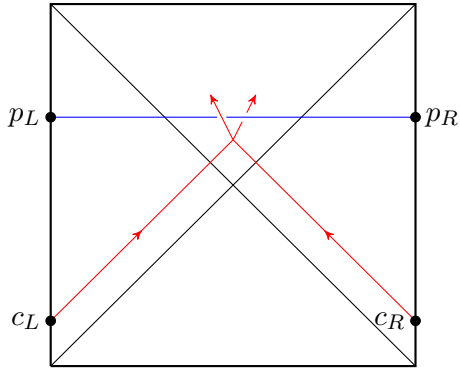
\begin{figure}
\centering
    \begin{tikzpicture}[scale=0.6]
    
    \draw[thick] (-4,-4) -- (-4,4) -- (4,4) -- (4,-4) -- (-4,-4);
    \draw (-4,-4) -- (4,4);
    \draw (-4,4) -- (4,-4);

    \draw[red,postaction={on each segment={mid arrow}}] (-4,-3) -- (0,1);
    \draw[red,postaction={on each segment={mid arrow}}] (4,-3) -- (0,1);
    \draw[red] (0,1) -- (0.2,1.4);
    \draw[red,->] (0.3,1.6) -- (0.5,2);
    \draw[blue] (-4,1.5) -- (-0.35,1.5);
    \draw[blue] (-0.15,1.5) -- (4,1.5);
    \draw[red,->] (0,1) -- (-0.5,2);

    \draw[black] plot [mark=*, mark size=3] coordinates{(-4,-3)};
    \node[left] at (-4,-3) {$c_L$};
    \draw[black] plot [mark=*, mark size=3] coordinates{(4,-3)};
    \node[left] at (4,-3) {$c_R$};
    
    \draw[black] plot [mark=*, mark size=3] coordinates{(4,1.5)};
    \node[left] at (-4,1.5) {$p_L$};
    \draw[black] plot [mark=*, mark size=3] coordinates{(-4,1.5)};
    \node[right] at (4,1.5) {$p_R$};
    
    \end{tikzpicture}
    \caption{An interior collision in the BTZ black hole. When light rays (red) from $c_L$, $c_R$ meet before the extremal surface (blue), we always have $I(V_L:V_R)=O(1/G_N)$, where $V_L=\hat{J}^+(c_L)\cap \hat{J}^-(p_L)$ and $V_R=\hat{J}^+(c_R)\cap \hat{J}^-(p_R)$. This gives a boundary signature for collisions behind the horizon, but only when the collisions are in the past of an extremal surface threading through the black hole.}\label{fig:interiorcollision}
\end{figure}

In the last section we focused on a setting where we take the entire left and right CFTs as input regions, and related interior collisions happening before the connected extremal surface to the circuit of figure \ref{fig:non-localcomputation}. 
In this section we also point out that a different choice of input regions allows us to find an interesting boundary signature of early interior collisions.\footnote{We thank Henry Lin who suggested applying the connected wedge theorem to the setting discussed here, and emphasized the significance to finding boundary signatures of interior collisions.} 
The basic conclusion of this section is summarized in figure \ref{fig:interiorcollision}. 

Let's return to the general construction of section \ref{sec:generaldblackholecase}, and consider inputting the infalling quantum system $A_L$, $A_R$ from points $c_L$, $c_R$ at the boundary. 
Recall that our construction involves cutting the left and right CFTs at points which we now denote $p_L$, $p_R$. 
In the quantum tasks language, we are taking the input regions to be points $c_L$ and $c_R$, and the output regions to be as before ($\mathcal{R}_-$ and $\mathcal{R}_+$ defined in equation \ref{eq:outputregions}).

With the inputs starting at a localized region in the boundary, we might expect only the entanglement in a geometrically local portion of the CFTs to be responsible for supporting the interior collision --- considering, for example, planar CFTs, entanglement between field degrees of freedom located far from the infalling systems seems unlikely to be necessary to support the interior collision. 
This suggests subregions of the two CFTs should be strongly correlated whenever there is an early collision in the interior. 
A natural candidate pair of subregions is
\begin{align}\label{eq:inputregions}
    V_L &= \hat{J}^+(c_L) \cap \hat{J}^-(p_L), \nonumber \\
    V_R &= \hat{J}^+(c_R) \cap \hat{J}^-(p_R).
\end{align}
These are the regions in the boundary with access to the inputs to the scattering process, and which can send signals to both of the output regions. 

The regions-based connected wedge theorem \cite{may2021holographic,may2022connected} would, applied to our setting, imply exactly that $V_L$ and $V_R$ develop $O(1/G_N)$ mutual information whenever $c_L$ and $c_R$ meet in the bulk in the past of the extremal surface $\gamma$ connecting $p_L$ and $p_R$. 
However, the connected wedge theorem is currently only proven in the context where all input and output regions are chosen from a single asymptotically AdS boundary, whereas our setting relies crucially on regions chosen from two disconnected asymptotic regions. 
Nonetheless, reviewing the geometrical argument for the connected wedge theorem at the level of \cite{may2020holographic} reveals it applies equally well to our setting. 
Less clear is adapting the more careful proof of \cite{may2022connected} here, but we leave this to future work.
As a check, we give an explicit calculation below showing this boundary entanglement signature for bulk collisions holds in a simple setting. 

\vspace{0.2cm}
\noindent \textbf{Planar BTZ black hole example}
\vspace{0.2cm}

We return to the two sided BTZ geometry
\begin{align}
    ds^2 = \frac{\ell^2}{\cos^2(w)} \left(-ds^2+dw^2+\cos^2(s)dX^2 \right).
\end{align}
We choose input points at $c_L$ and $c_R$ at $s=t_0$, $X=0$, $w=\pm \pi/2$ and a cut anchored to $p_L$, $p_R$ at $s=t_c$, $X=0$, $w=\pm \pi/2$. 
The extremal surface anchored to $p_L$ and $p_R$ sits at constant $s$.
Then, causally we have a scattering before this surface whenever
\begin{align}\label{eq:causalconditionBTZ}
    \boxed{t_c - t_0 \geq \pi/2}.
\end{align}
We define input regions according to equation \ref{eq:inputregions} and claim these regions have a connected entanglement wedge whenever we have a scattering in the bulk. 

To study this we first find the intervals defined by $V_L$ and $V_R$. 
Studying the forward and backward light cones of the input and output points, we find intersections at
\begin{align}
    f(t) &\equiv \frac{1+\tan(t/2)}{1-\tan(t/2)}, \nonumber \\
    X_m &= \pm \frac{1}{2} \ln \left( \frac{f(t_c)}{f(t_0)}\right) ,\nonumber \\
    t_m &= 2 \arctan \left(\frac{\sqrt{f(t_c)f(t_0)}-1}{\sqrt{f(t_c)f(t_0)}+1} \right).
\end{align}
We can adapt condition \ref{eq:periodic_minsurfacecondition} from section \ref{sec:BTZsolutionPeriodic} to investigate when intervals in two disconnected asymptotic regions have connected extremal surfaces that thread through the bulk. We find that we are in this connected phase when
\begin{align}
    \sinh\left(\Delta X/2\right) \geq \sec(T)
\end{align}
for $\Delta X$ the total width of the interval. 
Inserting $\Delta X=2|X_m|$ and $T=t_m$ here and simplifying we find exactly condition \ref{eq:causalconditionBTZ} above, so that meeting in the interior before the extremal surface always comes with a connected entanglement wedge, and hence large boundary mutual information. 
In fact, in this particular case a connected wedge also always happens with an interior collision, but this is not implied by the general reasoning above. 

\section{Finite dimensional model}\label{sec:finitedmodel}

We argued above that, so long as we don't induce a large backreaction, computations which happen in the region $J_T$ can be implemented as non-local computations, using the thermofield double state as the entangled resource system. 
Truncating the planar direction with ETW branes, or identifying it periodically, the entanglement as measured by the mutual information of the thermofield double is finite. 
However, the CFT Hilbert spaces are infinite dimensional. 
Results on non-local computation are typically proven in a finite dimensional setting. 
If we would like to apply them to holography then, we need to understand if the protocol carried out by the Alice's in the last section can be approximated by one involving only finite dimensional systems. 

To do this, a key ingredient is that in many holographic states the max entropy is close to the von Neumann entropy. 
Since the von Neumann entropy of our two CFTs is finite, this will let us show that the initial resource system, consisting of the entangled state of two CFTs, can be approximated closely by a finite dimensional system. 

\begin{figure}
    \centering
    \begin{tikzpicture}[scale=0.5]
    
    \draw[thick] (-5,-5) -- (-5,-3) -- (-3,-3) -- (-3,-5) -- (-5,-5);
    \node at (-4,-4) {$\mathbf{V}^L$};
    
    \draw[thick] (5,-5) -- (5,-3) -- (3,-3) -- (3,-5) -- (5,-5);
    \node at (4,-4) {$\mathbf{V}^R$};
    
    \draw[thick] (5,5) -- (5,3) -- (3,3) -- (3,5) -- (5,5);
    \node at (4,4) {$\mathbf{W}^R$};
    
    \draw[thick] (-5,5) -- (-5,3) -- (-3,3) -- (-3,5) -- (-5,5);
    \node at (-4,4) {$\mathbf{W}^L$};
    
    \draw[thick] (-4.5,-3) -- (-4.5,3);
    \node[below left] at (-4.5,-1.8) {$C_{L,L}$};
    
    \draw[thick] (4.5,-3) -- (4.5,3);
    \node[below right] at (4.5,-1.8) {$C_{R,R}$};
    
    \draw[thick] (-3.5,-3) to [out=90,in=-90] (3.5,3);
    \node[below right] at (-3.4,-1.9) {$C_{L,R}$};
    
    \draw[thick] (3.5,-3) to [out=90,in=-90] (-3.5,3);
    \node[below left] at (3.4,-1.9) {$C_{R,L}$};
    
    \draw[thick] (-3.5,-5) to [out=-90,in=-90] (3.5,-5);
    \draw[black] plot [mark=*, mark size=3] coordinates{(0,-7.05)};
    \node[below right] at (-3.4,-5) {$C_L$};
    \node[below left] at (3.4,-5) {$C_R$};
    
    \draw[thick] (-4.5,-6) -- (-4.5,-5);
    \node[below] at (-4.5,-6) {$A_L$};
    \draw[thick] (4.5,-6) -- (4.5,-5);
    \node[below] at (4.5,-6) {$A_R$};
    
    \draw[thick] (4.5,5) -- (4.5,6);
    \node[above] at (4.5,6) {$B_R$};
    
    \draw[thick] (-4.5,5) -- (-4.5,6);
    \node[above] at (-4.5,6) {$B_L$};
    
    \draw[thick] (3.5,5) -- (3.5,6);
    \draw[thick] (-3.5,5) -- (-3.5,6);
    
    \end{tikzpicture}
    \caption{General form of a non-local computation, with systems labelled. Note that $\mathbf{V}^L$, $\mathbf{V}^R$, $\mathbf{W}^L$, and $\mathbf{W}^R$ are isometries.}
    \label{fig:secondnon-local}
\end{figure}
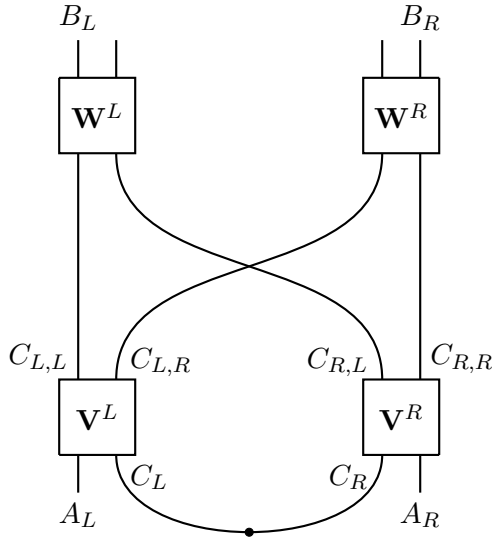

In more detail, we study the following general form of a non-local computation protocol, which we argue below captures the holographic setting.
\begin{definition}\label{def:generalNLQC} A \textbf{non-local computation protocol} takes the following form.
\begin{enumerate}
    \item The inputs are recorded into Hilbert spaces $\mathcal{H}_{A_L}$, $\mathcal{H}_{A_R}$, each consisting of $n$ qubits.  
    \item The resource system $\ket{\phi}_{C_LC_R}$ lives in a Hilbert space $\mathcal{H}_{C_{L}}\otimes \mathcal{H}_{C_{R}}$, where each tensor factor consists of $E$ qubits. 
    \item In the first round an isometry $\mathbf{V}^{L}_{A_LC_{L}\rightarrow C_{L,L}C_{L,R}}\otimes \mathbf{V}^{R}_{A_RC_{R}\rightarrow C_{R,L}C_{R,R}}$ is applied. Note that the combined system of $C_{L,L}$ and $C_{L,R}$ refers to the same Hilbert space as the combined system of $C_L$ and $A_L$ and similarly for the right systems. We write the systems this way after the first round of isometries to emphasize which part of the left and right get exchanged.
    \item In the second round isometries $\mathbf{W}^{L}_{C_{L,L}C_{R,L}\rightarrow B_LE_L}\otimes \mathbf{W}^R_{C_{L,R}C_{R,R}\rightarrow B_RE_R}$ are applied. The systems $E_L$ and $E_R$ are traced out. 
\end{enumerate}
\end{definition}
\noindent Notice that between the first and second rounds systems $C_{L,R}$ and $C_{R,L}$ have been exchanged. 
This interchange of systems corresponds to the communication round of the non-local computation. 
We illustrate this protocol in figure \ref{fig:secondnon-local}. 

We would like to justify modelling the protocol described in our black hole setting in this form, and in particular understand how large of Hilbert spaces are needed to capture the power of the holographic protocol. 
To do this, first note that the entropy of the left (or right) BCFT is
\begin{align}
    S(L) = \frac{A_{bh}}{4G_N} = 2\ell X^0
\end{align}
which in particular is finite.
Next recall that the smooth max entropy \cite{renner2005simple} is defined as
\begin{align}
    S_{max}^\epsilon(A)_\rho = \min_{\sigma_A:||\rho_A-\sigma_A||_1\leq \epsilon} \log (\rank \sigma_A)
\end{align}
and that the smooth max entropy and von Neumann entropy are close for holographic states \cite{bao2019beyond},
\begin{align}\label{eq:holographicSmax}
    S_{max}^{\epsilon}(A) = S(A) + a\frac{\log(1/\epsilon)}{\sqrt{G_N}}
\end{align}
where $a$ is independent of $G_N$, but can depend on the state $\rho$. 
This holds for holographic states as a consequence of the Renyi entropy taking the form,
\begin{align}\label{eq:genericSalpha}
    S_\alpha = \frac{s_\alpha}{G_N}
\end{align}
where $s_\alpha$ is independent of $G_N$, but is otherwise an arbitrary function of $\alpha$ and the choice of state. 
While \cite{bao2019beyond} does not explicitly consider holographic BCFT states, it is straightforward to see that \ref{eq:genericSalpha} continues to hold for these states, and hence we recover \ref{eq:holographicSmax} as well in our setting.\footnote{One can see this by noting that the cosmic brane prescription of \cite{dong2016gravity} applies to our setting, which yields a value of the Renyi entropy of the form \ref{eq:genericSalpha}.  
A comment is that in our setting the cosmic brane will anchor to the two ETW branes, so the Renyi entropy's are UV divergence free. 
We can also note that the form \ref{eq:genericSalpha} has been found directly from a CFT perspective in \cite{sully2021bcft}, though only for intervals ending on one CFT boundary (rather than anchored on both ends to CFT boundaries, as in our setting).} 

Let the state of the two BCFTs appearing in our thought experiment be $\ket{\Psi}_{C_LC_R}$.
Consider the reduced density matrix $\Psi_{C_L}$. 
Then as a consequence of \ref{eq:holographicSmax}, we have that there is a state $\Psi'_{C_L}$ with rank $S^\epsilon_{max}(C_L)$ which is $\epsilon$ close in trace distance to the state ${\Psi}_{C_L}$. 
Then using the Fuchs-Van-de-Graff inequality, 
\begin{align}\label{eq:Fuchs}
    1-\sqrt{F(\rho,\sigma)} \leq \frac{1}{2}||\rho-\sigma||_1 \leq \sqrt{1-F(\rho,\sigma)}
\end{align}
we have that $\sqrt{F}(\Psi_{C_L},\Psi_{C_L}') \geq 1-\epsilon/2$. 
Using Uhlmann's theorem \cite{uhlmann1976transition} we have
\begin{align}
    1-\epsilon/2 \leq \sqrt{F(\Psi_{C_L},\Psi_{C_L}')} = \max_{\Phi_{C_LC_R}} |\braket{\Phi}{\Psi}_{C_LC_R} |.
\end{align}
Thus there exists a purification of $\Psi_{C_L}'$, call it $\ket{\Psi'}_{C_LC_R}$, which has $|\braket{\Psi'}{\Psi}_{C_LC_R} | \geq 1-\epsilon/2$, or using the Fuchs van de Graff inequality again we have
\begin{align}
    || \ket{\Psi'}-\ket{\Psi}|| \leq \sqrt{\epsilon/2}.
\end{align}
Because this purifies $\Psi_{C_L}'$ we know $\ket{\Psi'}_{C_LC_R}$ has Schmidt rank $S_{max}^{\epsilon}(C_L)\equiv E$, so
\begin{align}
    \ket{\Psi'}_{C_LC_R} = \sum_{i=1}^{2^E} \sqrt{\lambda_i}\ket{i}_{C_L}\ket{i}_{C_R}.
\end{align}
Next, we define finite dimensional Hilbert spaces $\mathcal{H}_{C_L'}\otimes \mathcal{H}_{C_R'}$ whose basis vectors are identified with the Schmidt vectors of $\ket{\Psi'}_{C_LC_R}$. 
These each have dimension $2^E$. 

In our thought experiment, each of Alice$_L$ and Alice$_R$ act on their inputs plus their respective BCFTs to insert the inputs into the bulk. 
In our finite dimensional model, we should first locally prepare the BCFT states from the finite dimensional state $\ket{\Psi'}$.
This is possible because the finite dimensional entangled state shares the same Schmidt spectrum.
Each Alice then insert the inputs into the bulk. 
This is captured by the application of the first round operations $\mathbf{V}^L_{A_LC_L\rightarrow C_{L,L}C_{L,R}}\otimes \mathbf{V}^R_{A_RC_R\rightarrow C_{R,L}C_{R,R}}$. 
Next, the Alices divide the BCFTs into $X>0$ and $X<0$ portions, and communicate half of the degrees of freedom. 
Naively this step presents a challenge for our finite dimensional model to capture, since the entanglement across the $X=0$ cut in the CFT will be infinite, so we can't prepare states sharing approximately the same spectrum in a finite dimensional Hilbert space.  
However, a key feature of our finite dimensional model is that we allow isometries in the first round, and in particular systems $C_{L,L}$, $C_{L,R}$ and $C_{R,L}$, $C_{R,R}$ can be arbitrarily large. 
Thus we can choose the $X_+$ and $X_-$ regions to be separated by a small cut-off $\delta$, and our protocol captures the entanglement across this cut-off arbitrarily well. 
The dimension of the needed system will grow as $\delta \rightarrow 0$, but because we allow isometries, this is captured by our protocol above. 
Note that there is no dependence on the cut-off in the amount of entanglement used, which is fixed by the (finite) black hole area.  
The thought experiment then proceeds as expected with local interaction in the bulk, corresponding to swapping of subsystems on the boundary, and with the reconstruction operators that retrieve the outputs from the bulk, captured by the isometries $\mathbf{W}^{L}_{C_{L,L}C_{R,L}\rightarrow B_LE_L}\otimes \mathbf{W}^R_{C_{L,R}C_{R,R}\rightarrow B_RE_R}$.

\section{Computation in global AdS\texorpdfstring{$_{2+1}$}{TEXT}}\label{sec:AdS2+1computations}

In this section we study global AdS$_{2+1}$, and show that a similar claim to the two sided case holds: computation happening in the bulk of AdS can be reproduced as a non-local quantum computation, using entanglement controlled by the size of the scattering region. 

Towards showing this the main obstruction is that interactions in the bulk of global AdS$_{2+1}$ are most naturally related to augmented non-local computation, rather than non-local computation directly. 
This is related to the appearance of the ``side-regions'' that we define below, from which the extra systems discussed in figure \ref{fig:augmentednon-localcomputation} originate. 
To get around this, we approximate the original CFT with a pair of entangled BCFTs. 
The BCFTs live on the same geometry as the original CFT but with the side regions removed. 
We show that the dual of the BCFTs includes enough of the bulk geometry that they can still be used to perform the computation, and that they share similar entanglement to the relevant subregions of the original CFT.
Quantitatively, they share entanglement that becomes close to the area of the scattering region as the scattering region becomes large.\footnote{This justifies the assumption made in \cite{may2022complexity} that the entanglement between the inputs is responsible for supporting the computations happening inside the scattering region in the limit where the side regions are becoming small.
It also gives a quantitative bound on how much more entanglement can be necessary in the setting of large side regions.} 

\subsection{Bulk and boundary geometry of \texorpdfstring{$2\rightarrow 2$}{TEXT} task}\label{sec:bulkandboundarygeometry}

\begin{figure}
\begin{center}
\begin{subfigure}{0.45\textwidth}
\begin{center}
\tdplotsetmaincoords{10}{0}
    \begin{tikzpicture}[scale=1,tdplot_main_coords]
    \tdplotsetrotatedcoords{0}{20}{0}
    \draw (-2,0,0) -- (-2,4,0);
    \draw (2,0,0) -- (2,4,0);
    
    \begin{scope}[tdplot_rotated_coords]
    \begin{scope}[canvas is xz plane at y=0]
    \draw (0,0) circle [radius=2];
    \end{scope}
    
    \begin{scope}[canvas is xz plane at y=4]
    \draw (0,0) circle [radius=2];
    \end{scope}
    
    \draw[red,-triangle 45] (2,0,0) -- (0,2,0);
    \draw[red,-triangle 45] (-2,0,0) -- (0,2,0);
    \draw[red,-triangle 45] (0,2,0) -- (0,4,2);
    \draw[red,-triangle 45] (0,2,0) -- (0,4,-2);
    
    \draw plot [mark=*, mark size=2] coordinates{(2,0,0)};
    \node[below right] at (2,0,0) {$c_1$};
    
    \draw plot [mark=*, mark size=2] coordinates{(-2,0,0)};
    \node[below left] at (-2,0,0) {$c_2$};
    
    \draw plot [mark=*, mark size=1.5] coordinates{(0,2,0)};
    \node[left] at (0,2,0) {$p$};
    
    \draw plot [mark=*, mark size=2] coordinates{(0,4,-2)};
    \node[above left] at (0,4,-2) {$r_2$};
    
    \draw plot [mark=*, mark size=2] coordinates{(0,4,2)};
    \node[above right] at (0,4,2) {$r_1$};
    
    \end{scope}

\end{tikzpicture}
\end{center}
\caption{}
\label{fig:cylinder}
\end{subfigure}
\hfill
\begin{subfigure}{.45\textwidth}
\begin{center}
\begin{tikzpicture}[scale=1.2]

    \draw (-2,0) -- (2,0) -- (2,2) -- (-2,2) -- (-2,0);
    
    \draw[gray,fill=gray,opacity=0.25] (-2,2) -- (0,0) -- (2,2);
    
    \draw[gray,fill=gray,opacity=0.25] (-2,2) -- (-2,0) -- (0,2) -- (-2,2);
    
    \draw[gray,fill=gray,opacity=0.25] (2,2) -- (2,0) -- (0,2) -- (2,2);
    
    \draw[blue,fill=blue,opacity=0.25] (-1,2) -- (-2,1) -- (-2,0) -- (1,0) -- (-1,2);
    
    \draw[blue,fill=blue,opacity=0.25] (1,2) -- (2,1) -- (2,0) -- (-1,0) -- (1,2);
    
    \draw[blue,fill=blue,opacity=0.25] (2,1) -- (2,0) -- (1,0) -- (2,1);
    \draw[blue,fill=blue,opacity=0.25] (-2,1) -- (-2,0) -- (-1,0) -- (-2,1);
    
    \draw[black] plot [mark=*, mark size=2] coordinates{(-2,0)};
    \node[below left] at (-2,0) {$c_2$};
    \node at (0,0.5) {$\mathcal{V}_1$};
    
    \draw[black] plot [mark=*, mark size=2] coordinates{(2,0)};
    \node[below right] at (2,0) {$c_2$};
    \node at (1.8,0.5) {$\mathcal{V}_2$};
    
    \draw[black] plot [mark=*, mark size=2] coordinates{(0,0)};
    \node[below left] at (0,0) {$c_1$};
    
    \draw[black] plot [mark=*, mark size=2] coordinates{(-1,2)};
    \node[above right] at (-1,2) {$r_1$};
    
    \draw[black] plot [mark=*, mark size=2] coordinates{(1,2)};
    \node[above right] at (1,2) {$r_2$};
    
    \node at (0,-1) {$ $};
    
    \node at (1,0.5) {$\mathcal{X}_2$};
    \node at (-1,0.5) {$\mathcal{X}_1$};

\end{tikzpicture}
\end{center}
\caption{}
\label{fig:boundaryrectangle}
\end{subfigure}
\begin{subfigure}{.45\textwidth}
\begin{center}
\begin{tikzpicture}[scale=0.8]

\draw[mid arrow, red] (0,0) -- (2,2);
    \draw[mid arrow, red] (0,0) -- (-2,2);
    \draw[mid arrow, red] (-2,-2) -- (0,0);
    \draw[mid arrow, red] (2,-2) -- (0,0);
    
    \draw plot [mark=*, mark size=2] coordinates{(0,0)};
    \draw plot [mark=*, mark size=2] coordinates{(-2,-2)};
    \draw plot [mark=*, mark size=2] coordinates{(2,2)};
    \draw plot [mark=*, mark size=2] coordinates{(-2,2)};
    \draw plot [mark=*, mark size=2] coordinates{(2,-2)};

\end{tikzpicture}

\end{center}
\caption{}
\label{fig:bulknetwork}
\end{subfigure}
\hfill
\begin{subfigure}{.45\textwidth}
\begin{center}
\begin{tikzpicture}[scale=0.8]

    \draw[mid arrow, red] (-2,-2) -- (0,0);
    \draw[red, mid arrow] (0,0) -- (2,2);
    \draw[mid arrow, red] (2,-2) -- (0.1,-0.1);
    \draw[mid arrow, red] (-0.1,0.1) -- (-2,2);
    \draw[mid arrow, red] (-2,-2) -- (-2,2);
    \draw[mid arrow, red] (2,-2) -- (2,2);
    
    \draw[mid arrow, red] (-3.5,0) -- (-2,2);
    \draw[mid arrow, red] (3.5,0) -- (2,2);
    
    \draw plot [mark=*, mark size=2] coordinates{(-2,-2)};
    \draw plot [mark=*, mark size=2] coordinates{(2,2)};
    \draw plot [mark=*, mark size=2] coordinates{(-2,2)};
    \draw plot [mark=*, mark size=2] coordinates{(2,-2)};

    \draw plot [mark=*,mark size=2] coordinates{(-3.5,0)};
    \draw plot [mark=*,mark size=2] coordinates{(3.5,0)};
    
    \node[below] at (-3.5,0) {$\mathcal{X}_1$};
    \node[below] at (3.5,0) {$\mathcal{X}_2$};
    
    \node[below] at (-2,-2) {$\mathcal{V}_1$};
    \node[below] at (2,-2) {$\mathcal{V}_2$};

\end{tikzpicture}

\end{center}
\caption{}
\label{fig:boundarynetwork}
\end{subfigure}
\caption{\textbf{(a)} Global AdS$_{2+1}$ with an example choice of input and output points. These points have a non-empty scattering region in the bulk, but an empty scattering region in the boundary. \textbf{(b)} Boundary view of AdS$_{2+1}$ with an illustration of the input regions $\mathcal{V}_1,\mathcal{V}_2$ and side regions $\mathcal{X}_1, \mathcal{X}_2$. \textbf{(c)} Causal structure present in the bulk of AdS, with the choice of input and output regions shown in figure (a). \textbf{(d)} Causal structure present in the boundary of AdS with the same choice of input and output regions.}
\end{center}
\end{figure}
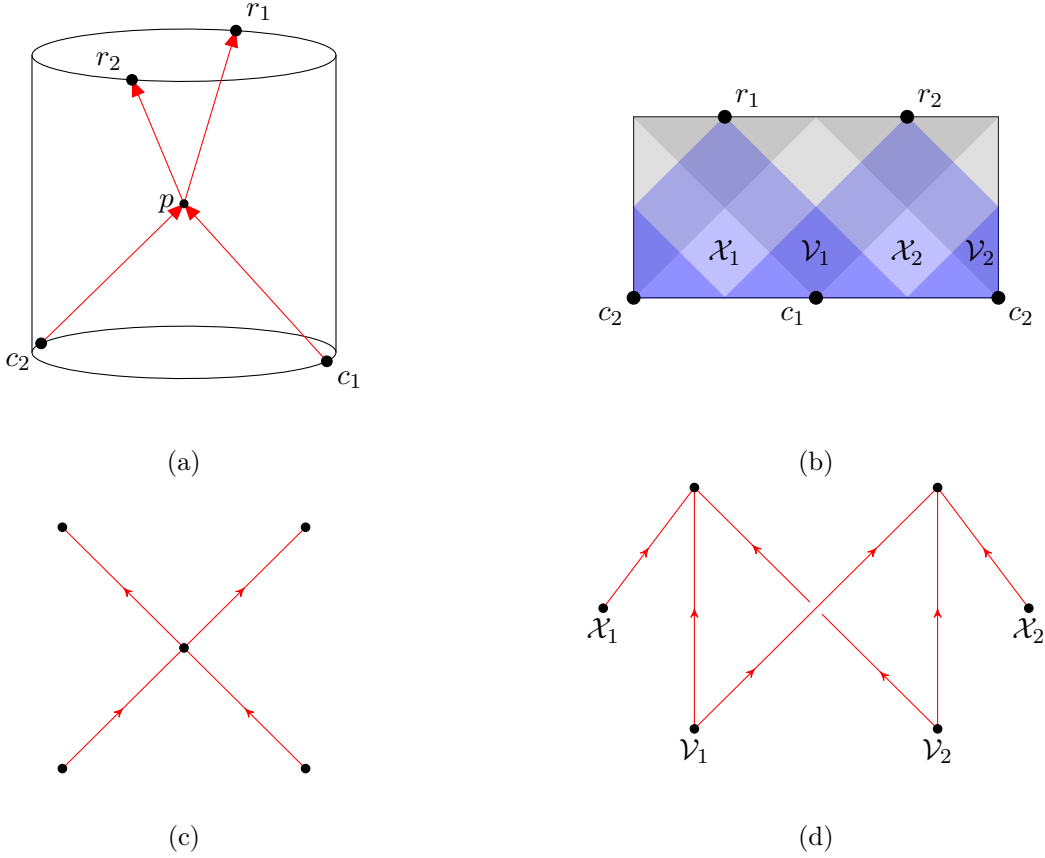

In this section we give the geometrical set-up of input and output points at the boundary of AdS$_{2+1}$. 
To begin, consider the picture in figure \ref{fig:cylinder}, showing an AdS$_{2+1}$ spacetime.
In figure \ref{fig:boundaryrectangle} we show the dual boundary spacetime.
The basic observation is that, choosing four points in the boundary, we can have it happen that one can meet in the bulk but not in the boundary. 
To discuss this more carefully, define the bulk \emph{scattering region},
\begin{align}
    J[c_1,c_2\rightarrow r_1,r_2] \equiv J^+(c_1) \cap J^+(c_2) \cap J^-(r_1) \cap J^-(r_2).
\end{align}
The observation is that we can have geometries and choices of points $c_1,c_2,r_1,r_2$ such that $J[c_1,c_2\rightarrow r_1,r_2]$ is non-empty, while the object $\hat{J}[c_1,c_2\rightarrow r_1,r_2]$ obtained by replacing each bulk light cone with its boundary restriction is empty. 

We introduce some definitions to capture relevant aspects of the boundary geometry. 
Define the \emph{input regions}
\begin{align}
    \mathcal{V}_1 &= \hat{J}^+(c_1) \cap \hat{J}^-(r_1) \cap \hat{J}^-(r_2),\\
    \mathcal{V}_2 &= \hat{J}^+(c_2) \cap \hat{J}^-(r_1) \cap \hat{J}^-(r_2).
\end{align}
Further, define the \emph{side regions},
\begin{align}
    \mathcal{X}_1 &= \hat{J}^-(r_1) \cap \overline{[\mathcal{V}_1\cup \mathcal{V}_2]},\\
    \mathcal{X}_2 &= \hat{J}^-(r_2) \cap \overline{[\mathcal{V}_1\cup \mathcal{V}_2]},
\end{align}
where $\bar{\mathcal{A}}$ denotes the spacelike complement of spacetime region $\mathcal{A}$. 
See figure \ref{fig:boundaryrectangle} for an illustration of these regions. 

Figure \ref{fig:bulknetwork} shows the causal features of the bulk schematically, in the form of a causal network.
The central vertex represents the (non-empty) scattering region. 
In figure \ref{fig:boundarynetwork}, we show the causal network describing the boundary. 
Notice that in the boundary, we don't have exactly the usual network describing a non-local quantum computation. 
This is because of the regions $\mathcal{X}_1$ and $\mathcal{X}_2$ that sit between the regions $\mathcal{V}_1$ and $\mathcal{V}_2$. 
We call this the \emph{augmented NLQC} scenario. 
This complicates the discussion, as it is only correlation between $\mathcal{V}_1$ and $\mathcal{V}_2$ that we will have control over, while e.g. $I(V_1X_1:V_2X_2)$ is divergent.  

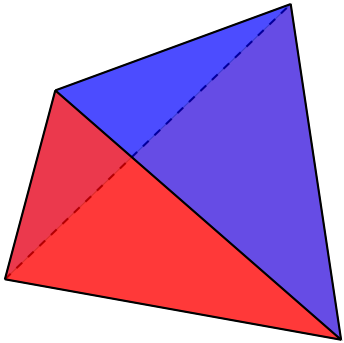
\begin{figure}
\centering
\tdplotsetmaincoords{15}{0}
\begin{tikzpicture}[scale=0.9,tdplot_main_coords]
\tdplotsetrotatedcoords{0}{35}{0}
    
    \begin{scope}[tdplot_rotated_coords]
    
    \draw[thick,dashed] (-3,0,0) -- (0,4,3);
    
    \draw[opacity=0.25,fill=red] (0,4,3) -- (-3,0,0) -- (3,0,0) -- (0,4,3);
    \draw[opacity=0.25,fill=blue] (-3,0,0) -- (0,4,3) -- (0,4,-3) -- (-3,0,0);
    \draw[opacity=0.6,fill=blue] (3,0,0) -- (0,4,3) -- (0,4,-3) -- (3,0,0);
    \draw[opacity=0.7,fill=red] (0,4,-3) -- (-3,0,0) -- (3,0,0) -- (0,4,-3);
    
    \draw[thick] (-3,0,0) -- (3,0,0);
    \draw[thick] (0,4,-3) -- (0,4,3);
    
    \draw[thick] (-3,0,0) -- (0,4,-3);
    
    \draw[thick] (3,0,0) -- (0,4,-3);
    \draw[thick] (3,0,0) -- (0,4,3);
    
    \end{scope}
    \end{tikzpicture}
    \caption{A scattering region in AdS$_{2+1}$. The lower edge is the ridge, $r$.}
    \label{fig:scatteringregion}
\end{figure}

We recall some further geometric facts established in \cite{may2020holographic}. 
First, an interesting observation was made that the scattering region sits inside of the entanglement wedge of $\mathcal{V}_1\cup \mathcal{V}_2$. 
In notation,
\begin{align}
    J[c_1,c_2\rightarrow r_1,r_2] \subseteq E_{\mathcal{V}_1\mathcal{V}_2}. 
\end{align}
Second, we will measure the size of the scattering region in terms of the area of its lower edge, which we call the \emph{ridge}, defined by
\begin{align}
    r = \partial J^+(c_1)\cap \partial J^+(c_2) \cap J^-(r_1) \cap J^-(r_2).
\end{align}
We illustrate the scattering region and the placement of the ridge in figure \ref{fig:scatteringregion}. 
For the CFT vacuum state, \cite{may2020holographic} showed that the area of this surface has a simple boundary expression,
\begin{align}
    I(V_1:V_2)_\Psi = \frac{\text{area}(r)}{4G_N}.
\end{align}
This becomes $\geq$ away from the vacuum. 

Our goal below will be to relate computations happening inside the scattering region to (non-augmented) non-local computations. 
This connection was argued for based on the heuristic of ignoring the side regions in \cite{may2019quantum,may2020holographic}, but here we address this more carefully. 
Following our discussion of the two sided black hole case, our strategy is to set boundary conditions at the edge of the $\mathcal{V}_1$, $\mathcal{V}_2$ regions, and remove the side region degrees of freedom from our system. 
This replaces our initial CFT with two, entangled BCFTs that (in a sense we make precise) approximate the original CFT, and in particular share much of the original bulk geometry. 
In fact, we will see that it shares enough of the original geometry so as to still support the same bulk computations. 

\subsection{ETW brane solutions}

\begin{figure}
    \centering
    \begin{subfigure}{.45\textwidth}
    \includegraphics[scale=0.45]{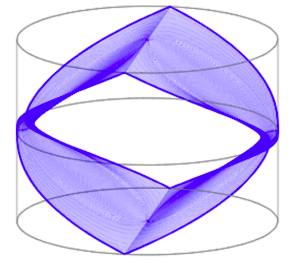}
    \caption{}
    \label{fig:twobranes}
    \end{subfigure}
    \caption{Lorentzian global AdS$_{2+1}$ showing a solution with two ETW branes. These solutions can be prepared via the Euclidean path integral, as studied in appendix \ref{appendix:global2+1}. The solution shown is the minimal action one when the intervals cut out by the brane have angular radius below $24^\circ$.}
    \label{fig:branefromPI}
\end{figure}

The particular solutions we need can be prepared via a Euclidean path integral.
Recall that pure global AdS$_{2+1}$ is prepared by the Euclidean path integral on the infinite cylinder. 
After a Wick rotation, and using global coordinates, the resulting solution is described by the metric
\begin{align}\label{eq:pureAdSglobal}
    ds^2= \frac{\ell^2}{\cos^2(r)}\left( -dt^2 + dr^2 + \sin^2(r)d\phi^2 \right).
\end{align}
We modify the Euclidean path integral by cutting out two (roughly) disk shaped regions, centered at $t=0$ and located antipodally. 
In appendix \ref{appendix:global2+1} we solve for the brane solution explicitly.\footnote{These brane solutions were also obtained in \cite{may2021interpolating}, although there multi-brane solutions are obtained by sewing together single brane Lorentzian solutions, and no explicit path integral giving the multi-brane solution is given. We prepare the two brane solution to allow studying the transition between a connected and disconnected geometry.} 
For $T=0$, we find the trajectory of each brane is described by
\begin{align}\label{eq:branetrajectory}
    \cos(t-t_0) = \frac{1}{\cos(\Delta \phi)} \cos(\phi-\phi_0) \sin r
\end{align}
and the interior geometry is a portion of the pure AdS geometry described by equation \ref{eq:pureAdSglobal}. 
We construct these solutions in detail in appendix \ref{appendix:global2+1}. 
Similar solutions for general values of $T$ exist. 
The parameter $\Delta \phi$ is the angular radius of the brane at $t=t_0$.
We can intersect the interior regions of several branes, placed at different $\phi_0$ and potentially with different widths, to construct multibrane solutions. 

Note that if we make the BCFT intervals too small, a new solution will dominate the path integral where the branes connect in the opposite way, so that the two BCFTs are not connected through the bulk geometry. 
We will assume we are in the configuration shown, with a connected geometry.
In appendix \ref{appendix:global2+1}, we find that we are in the connected geometry whenever the branes occupy intervals with angular radius $\Delta \phi \lesssim 24^\circ$.

An important view on this brane is its intersection with the $t=\pi/4$ slice.
This is given by 
\begin{align}
    1 = \frac{1}{\cos(\Delta \phi)} \cos(\phi-\phi_0) \sin(r).
\end{align}
We can observe that this is the same trajectory as the RT surface anchored to the same endpoints.\footnote{Compare e.g. to 6.1.26 in \cite{rangamani2017holographic}, after the coordinate transformation $\rho= \tan r$. This is true for zero tension branes and when comparing to extremal surfaces in pure AdS, but not in general.}
Further, we can observe that the brane sits inside the causal future of this $t=t_0$ surface --- this follows from a simple calculation showing that the constant $\theta$ curves, which foliate the full brane, are everywhere timelike or null. 

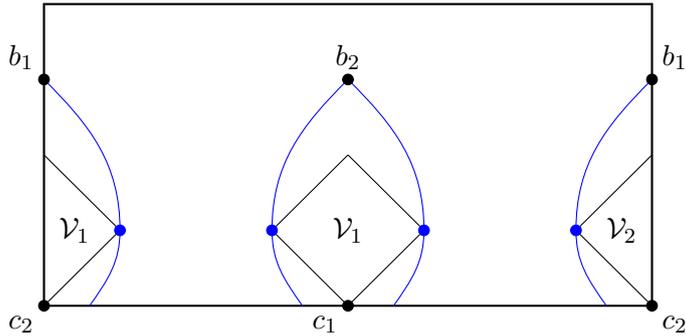
\begin{figure}
\centering
\begin{tikzpicture}[scale=2]

    \draw[thick] (-2,0) -- (2,0) -- (2,2) -- (-2,2) -- (-2,0);
    \draw (-0.5,0.5) -- (0,0) -- (0.5,0.5);
    \draw (-2,0) -- (-1.5,0.5);
    \draw (-2,1) -- (-1.5,0.5);
    \draw (0.5,0.5) -- (0,1);
    
    \draw (-0.5,0.5) -- (0,1);
    \draw (2,0) -- (1.5,0.5) -- (2,1);
    
    \draw[blue] (-0.5,0.5) to [out=90,in=-135] (0,1.5);
    \draw[blue] (-1.5,0.5) to [out=90,in=-45] (-2,1.5);
    
    \draw[blue] plot [mark=*, mark size=1] coordinates{(-0.5,0.5)};
    \draw[blue] plot [mark=*, mark size=1] coordinates{(-1.5,0.5)};
    
    \draw[blue] (0.5,0.5) to [out=90,in=-45] (0,1.5);
    \draw[blue] (1.5,0.5) to [out=90,in=-135] (2,1.5);
    
    \draw[blue] plot [mark=*, mark size=1] coordinates{(0.5,0.5)};
    \draw[blue] plot [mark=*, mark size=1] coordinates{(1.5,0.5)};
    
    \draw[blue] (0.5,0.5) to [out=-90,in=55] (0.3,0);
    \draw[blue] (1.5,0.5) to [out=-90,in=125] (1.7,0);
    
    \draw[blue] (-0.5,0.5) to [out=-90,in=125] (-0.3,0);
    \draw[blue] (-1.5,0.5) to [out=-90,in=55] (-1.7,0);
    
    \draw[black] plot [mark=*, mark size=1] coordinates{(-2,0)};
    \node[below left] at (-2,0) {$c_2$};
    \node at (0,0.5) {$\mathcal{V}_1$};
    
    \node at (-1.8,0.5) {$\mathcal{V}_1$};
    \node at (1.8,0.5) {$\mathcal{V}_2$};
    
    \draw[black] plot [mark=*, mark size=1] coordinates{(2,0)};
    \node[below right] at (2,0) {$c_2$};
    
    \draw[black] plot [mark=*, mark size=1] coordinates{(0,0)};
    \node[below left] at (0,0) {$c_1$};
    
    \node at (0,-1) {$ $};
    
    \draw[black] plot [mark=*, mark size=1] coordinates{(0,1.5)};
    \node[above] at (0,1.5) {$b_2$};
    \draw[black] plot [mark=*, mark size=1] coordinates{(-2,1.5)};
    \node[above left] at (-2,1.5) {$b_1$};
    \draw[black] plot [mark=*, mark size=1] coordinates{(2,1.5)};
    \node[above right] at (2,1.5) {$b_1$};
    
\end{tikzpicture}
\caption{Boundary picture, where the $\mathcal{X}_i$ regions have been removed by replacing them with boundary conditions on the $\mathcal{V}_i$ regions. The brane trajectories are shown in blue. Note that the region behind the branes is empty --- no spacetime or degrees of freedom are associated with it. These two boundary spacetime regions are connected through the bulk geometry.}
\label{fig:boundarybranetrajectories}
\end{figure}

The trajectory of the branes endpoint on the boundary is found by setting $r=\pi/2$ in \ref{eq:branetrajectory}, which gives
\begin{align}
    \cos(t-t_0) = \frac{1}{\cos(\Delta \theta)} \cos(\phi-\phi_0).
\end{align}
We are interested specifically in a setting with two branes, one centered at $t=\pi/4$, $\phi=\pi/2$, the other at $t=\pi/4,\theta=\pi/2$.
We find that these intersect at 
\begin{align}
    b_1=(t=3\pi/4,\phi=0),\\
    b_2=(t=3\pi/4,\phi=\pi).
\end{align}
We show the boundary brane trajectories in figure \ref{fig:boundarybranetrajectories}. 

These BCFT geometries were studied in \cite{van2020spacetime,may2021interpolating}, where it was pointed out that they approximate the original CFT state in a precise sense. 
In particular, as the radius of the circles on which boundary conditions have been set goes to zero, the reduced density matrix on the remaining degrees of freedom approaches the density matrix of the corresponding subsystems of the CFT. 
Because these BCFT states approximate the original CFT state while having the side regions removed, they are natural candidates to use to connect computation in the bulk to (un-augmented) non-local computation. 
We take this up in the next section. 

\subsection{Non-local computation and the scattering region}

In this section we give a protocol that allows whatever computations that can happen in the scattering region to be performed non-locally, using a resource system with mutual information controlled by the size of the scattering region.
This holds so long as the computation does not induce too large of a backreaction, and the angular radius of the extra regions is smaller than the threshold needed to obtain a connected brane geometry, $\Delta \phi \lesssim 24^\circ$.

To begin, suppose some quantum task $T$ can be completed in the bulk, with some particular placement of input and output regions. 
This placement of points defines spacetime regions $\mathcal{V}_1$, $\mathcal{V}_2$ in the boundary.
We take the background state to be the CFT vacuum, corresponding to pure AdS in the bulk.  
The bulk picture has input systems $A_1$, $A_2$ meet in the scattering region, interact, then exit the scattering region and travel towards the output points. 
This can be arranged by acting with unitaries $\mathbf{V}_1$, $\mathbf{V}_2$ on regions $\mathcal{V}_1, \mathcal{V}_2$ to insert the inputs and computing device into the bulk, then allowing the CFT to time evolve. 

From this process, we build a non-local computation protocol that completes the same task, in the non-augmented scenario. 
Our strategy is to use $\ket{\tilde{\Psi}}_{{V}_1{V}_2}$ as the resource system, which we define as the dual state to the two brane solution discussed in the last section. 
We choose the size and placement of the branes such that the side regions $\mathcal{X}_i$ are cut out, as shown in figure \ref{fig:boundarybranetrajectories}. 
We denote the spacetime dual to $\ket{\Psi}_{{V}_1{V}_2{X}_1{X}_2}$ as $\mathcal{M}$, and the spacetime dual to $\ket{\tilde{\Psi}}_{{V}_1{V}_2}$ as $\tilde{\mathcal{M}}$.
It will be convenient to view $\tilde{\mathcal{M}}$ as a subset of $\mathcal{M}$, whose boundary is defined by the brane trajectory. 

To describe the protocol, we will adopt the operational language briefly used in section \ref{sec:generaldblackholecase}, and consider Alice$_L$ and Alice$_R$, who will each initially hold one of the BCFTs. 
To construct the protocol, first observe that $\tilde{\mathcal{M}}$ contains the scattering region. 
To see this, notice that the entangling surface $\gamma_{\mathcal{V}_1\mathcal{V}_2}$ in the original geometry sits exactly on the two branes, along the moment of time symmetry. 
Because the branes follow timelike trajectories, they sit outside the domain of dependence of $E_{\mathcal{V}_1\mathcal{V}_2}$. 
Finally, recall from section \ref{sec:bulkandboundarygeometry} that the scattering region sits inside of the entanglement wedge $E_{\mathcal{V}_1\mathcal{V}_2}$, so that it also sits inside of $\tilde{\mathcal{M}}$.

\begin{figure}
\centering
\begin{tikzpicture}[scale=2]

    \draw[thick] (-2,0) -- (2,0) -- (2,2) -- (-2,2) -- (-2,0);
    \draw (-0.5,0.5) -- (0,0) -- (0.5,0.5);
    \draw (-2,0) -- (-1.5,0.5);
    \draw[very thick,red] (-2,1) -- (-1.5,0.5);
    \draw[very thick,orange] (0.5,0.5) -- (0,1);
    
    \draw[very thick,red] (-0.5,0.5) -- (0,1);
    \draw (2,0) -- (1.5,0.5);
    \draw[very thick, orange] (1.5,0.5) -- (2,1);
    
    \draw[blue] (0.5,0.5) to [out=90,in=-45] (0,1.5);
    \draw[blue] (1.5,0.5) to [out=90,in=-135] (2,1.5);
    
    \draw[blue] (-0.5,0.5) to [out=90,in=-135] (0,1.5);
    \draw[blue] (-1.5,0.5) to [out=90,in=-45] (-2,1.5);

    \draw[blue] (0.5,0.5) to [out=-90,in=55] (0.3,0);
    \draw[blue] (1.5,0.5) to [out=-90,in=125] (1.7,0);
    
    \draw[blue] (-0.5,0.5) to [out=-90,in=125] (-0.3,0);
    \draw[blue] (-1.5,0.5) to [out=-90,in=55] (-1.7,0);
    
    \draw[black] plot [mark=*, mark size=1] coordinates{(-2,0)};
    \node[below left] at (-2,0) {$c_2$};
    \node at (0,0.5) {$\mathcal{V}_1$};
    
    \node at (-1.8,0.5) {$\mathcal{V}_1$};
    \node at (1.8,0.5) {$\mathcal{V}_2$};
    
    \draw[black] plot [mark=*, mark size=1] coordinates{(2,0)};
    \node[below right] at (2,0) {$c_2$};
    
    \draw[black] plot [mark=*, mark size=1] coordinates{(0,0)};
    \node[below left] at (0,0) {$c_1$};
    
    \node at (0,-1) {$ $};
    
    \draw[black] plot [mark=*, mark size=1] coordinates{(0,1.5)};
    \node[above] at (0,1.5) {$b_2$};
    \draw[black] plot [mark=*, mark size=1] coordinates{(-2,1.5)};
    \node[above left] at (-2,1.5) {$b_1$};
    \draw[black] plot [mark=*, mark size=1] coordinates{(2,1.5)};
    \node[above right] at (2,1.5) {$b_1$};
    
    \draw[red] plot [mark=*, mark size=1] coordinates{(-0.5,0.5)};
    \draw[red] plot [mark=*, mark size=1] coordinates{(-1.5,0.5)};
    \draw[orange] plot [mark=*, mark size=1] coordinates{(0.5,0.5)};
    \draw[orange] plot [mark=*, mark size=1] coordinates{(1.5,0.5)};
    
\end{tikzpicture}
\caption{In the communication round of the protocol, degrees of freedom on the left future boundary of $\mathcal{V}_1$ and right future boundary of $\mathcal{V}_2$ are passed to Alice$_1$, these are shown in red. Degrees of freedom on the right future boundary of $\mathcal{V}_1$ and left future boundary of $\mathcal{V}_2$ are passed to Alice$_2$, these are shown in orange.}
\label{fig:boundarysystemdistribution}
\end{figure}
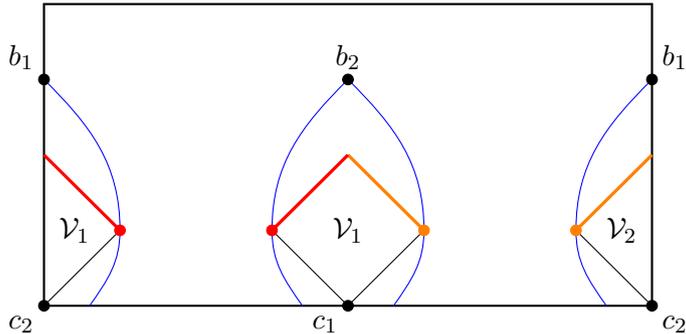

The first round of the protocol then is to take the state $\ket{\tilde{\Psi}}_{{V}_1{V}_2}$ and act on it with $\mathbf{V}_1 \otimes \mathbf{V}_2$ --- the same unitaries as in the original state $\Psi$.
In the bulk picture, the same computation happens inside of the scattering region as in the original geometry, and the outputs begin moving towards the output locations. 
In the communication round, Alice$_1$ and Alice$_2$ redistribute subsystems of the CFT according to the picture in figure \ref{fig:boundarysystemdistribution}.
Label the systems held by Alice$_i$ after the communication round by $W_i$, and the corresponding subregions by $\mathcal{W}_i$. 

\begin{figure*}
    \centering
    \begin{subfigure}{0.45\textwidth}
    \centering
    \begin{tikzpicture}[scale=0.8]

    \node[right] at (0,1) {$\gamma_{\mathcal{W}_1}'$};
    
    \draw [domain=-45:45,black,dashed] plot ({3*cos(\x)},{3*sin(\x)});
    \draw [domain=45:135,black] plot ({3*cos(\x)},{3*sin(\x)});
    \draw [domain=135:225,black,dashed] plot ({3*cos(\x)},{3*sin(\x)});
    \draw [domain=225:315,black] plot ({3*cos(\x)},{3*sin(\x)});

    \fill[red,opacity=0.25,variable=\x,domain=0:45] plot ({3*sin(\x-180)}, {3*cos(\x-180)}) -- (-2.12, -2.12) to [out=45,in=-45] (-2.12, 2.12) -- plot ({3*sin(\x-45)}, {3*cos(\x-45)});

    \draw[ultra thick,red,variable=\x,domain=0:45] plot ({3*sin(\x-180)}, {3*cos(\x-180)});
    \draw[ultra thick,red,variable=\x,domain=0:45] plot ({3*sin(\x-45)}, {3*cos(\x-45)});

    \draw[ultra thick,orange,variable=\x,domain=0:-45] plot ({3*sin(\x-180)}, {3*cos(\x-180)});
    \draw[ultra thick,orange,variable=\x,domain=0:45] plot ({3*sin(\x)}, {3*cos(\x)});

    \node[right] at (1.2,0) {$b_2$};
    \node[left] at (-1.2,0) {$b_1$};

    \draw[thick,blue] (0,3) -- (0,-3);
    
    \draw[gray, thick] (2.12, 2.12) to [out=-135,in=135] (2.12, -2.12);
    \draw[gray, thick] (-2.12, 2.12) to [out=-45,in=45] (-2.12, -2.12);
    
    \node[above right] at ({3*cos(75)},{3*sin(75)}) {$\mathcal{W}_2$};
    \node[above left] at ({3*cos(105)},{3*sin(105)}) {$\mathcal{W}_1$};

    \node[below right] at ({3*cos(-75)},{3*sin(-75)}) {$\mathcal{W}_2$};
    \node[below left] at ({3*cos(-105)},{3*sin(-105)}) {$\mathcal{W}_1$};
        
    \end{tikzpicture}
    \caption{}
    \label{fig:threadedsurfaceglobal}
    \end{subfigure}
    \hfill
    \begin{subfigure}{0.45\textwidth}
    \centering
    \begin{tikzpicture}[scale=0.8]

    \fill[red,opacity=0.25,variable=\x,domain=0:45] plot ({3*sin(\x-180)}, {3*cos(\x-180)}) -- (-2.12, -2.12) to [out=45,in=-110] (-1.4, -1) to [out=-30,in=90] (0,-3);

    \fill[red,opacity=0.25,variable=\x,domain=0:45] plot ({3*sin(\x-45)}, {3*cos(\x-45)}) -- (-2.12, 2.12) to [out=-45,in=110] (-1.4, 1) to [out=30,in=-90] (0,3);
 
    \draw [domain=-45:45,black,dashed] plot ({3*cos(\x)},{3*sin(\x)});
    \draw [domain=45:135,black] plot ({3*cos(\x)},{3*sin(\x)});
    \draw [domain=135:225,black,dashed] plot ({3*cos(\x)},{3*sin(\x)});
    \draw [domain=225:315,black] plot ({3*cos(\x)},{3*sin(\x)});

    \draw[thick,blue] (0,3) to [out=-90,in=30] (-1.4,1);
    \node at (0,0) {$\gamma_{\mathcal{W}_1}''$};
    \draw[->,gray] (0,0.5) to[out=90,in=-45] (-0.5,1.5);
    \draw[->,gray] (0,-0.5) to[out=-90,in=45] (-0.5,-1.5);
    \draw[thick,blue] (0,-3) to [out=90,in=-30] (-1.4,-1);

    \node[right] at (1.2,0) {$b_2$};
    \node[left] at (-1.2,0) {$b_1$};
    
    \draw[gray, thick] (2.12, 2.12) to [out=-135,in=135] (2.12, -2.12);
    \draw[gray, thick] (-2.12, 2.12) to [out=-45,in=45] (-2.12, -2.12);
    
    \node[above right] at ({3*cos(75)},{3*sin(75)}) {$\mathcal{W}_2$};
    \node[above left] at ({3*cos(105)},{3*sin(105)}) {$\mathcal{W}_1$};

    \node[below right] at ({3*cos(-75)},{3*sin(-75)}) {$\mathcal{W}_2$};
    \node[below left] at ({3*cos(-105)},{3*sin(-105)}) {$\mathcal{W}_1$};

    \draw[ultra thick,red,variable=\x,domain=0:45] plot ({3*sin(\x-180)}, {3*cos(\x-180)});
    \draw[ultra thick,red,variable=\x,domain=0:45] plot ({3*sin(\x-45)}, {3*cos(\x-45)});

    \draw[ultra thick,orange,variable=\x,domain=0:-45] plot ({3*sin(\x-180)}, {3*cos(\x-180)});
    \draw[ultra thick,orange,variable=\x,domain=0:45] plot ({3*sin(\x)}, {3*cos(\x)});
        
    \end{tikzpicture}
    \caption{}
    \label{fig:braneanchoredsurfaceglobal}
    \end{subfigure}
    \caption{Connected (a) and disconnected (b) configurations of the minimal surface enclosing the red system shown in figure \ref{fig:boundarysystemdistribution}. The orange subsystem has a similar transition and entanglement wedge.}
    \label{fig:rtsurfacesandbranes}
\end{figure*}
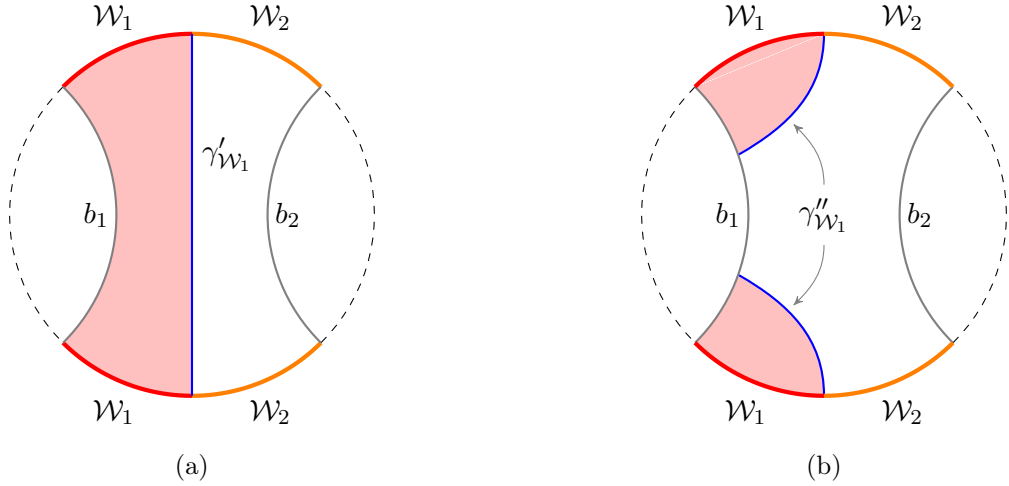

It remains to understand if, in the second round, the output systems $B_i$ can be recovered by Alice$_i$. 
To understand this, we should ask where the entanglement wedges of the $\mathcal{W}_i$ subregions sit.
See figure \ref{fig:rtsurfacesandbranes} for an illustration of the surfaces appearing in this paragraph. 
Considering $\gamma_{\mathcal{W}_1}$, there are two candidate geodesics for this surface: the line $\gamma_{\mathcal{W}_1}'$ at constant time connecting these two points straight through the bulk, or a surface that attaches to the brane, call it $\gamma_{\mathcal{W}_1}''$.
Taking the brane radius $R$ sufficiently small, we expect $\gamma_{\mathcal{W}_1}'$ is minimal.
Explicitly, in appendix \ref{appendix:global2+1} we find that this happens whenever the angular radius of the extra regions $\Delta \phi$ is less than $39^\circ$.
Since this is larger than the condition we already took to get a connected geometry, this adds no new constraint. 

\begin{figure}
    \centering
    \begin{tikzpicture}

        \fill[red,opacity=0.25] (1,2) to[out=90,in=-135] (1.7,3.3) -- (2.5,2.5) -- (1.2,1.2) to[out=120,in=-90] (1,2);
        \fill[orange,opacity=0.25] (4,2) to[out=90,in=-45] (3.3,3.3) -- (2.5,2.5) -- (3.8,1.2) to[out=60,in=-90] (4,2);

        \draw[thick,fill=blue,opacity=0.5] (2,2) -- (3,2) -- (2.5,2.5) -- cycle;
        
        \draw[thick] (0,0) -- (0,5);
        \draw[thick] (5,0) -- (5,5);
        \draw[thick] (0,0) -- (5,5);
        \draw[thick] (5,0) -- (0,5);

        \draw[black] plot [mark=*, mark size=2] coordinates{(0,5)};
        \node[left] at (0,5) {$r_1$};
        \draw[black] plot [mark=*, mark size=2] coordinates{(5,5)};
        \node[right] at (5,5) {$r_2$};

        \node[above] at (2.5,2.75) {$\gamma_{\mathcal{W}_i}$};

        \draw[gray,thick] (2.5,0) to[out=135,in=-90] (1,2) to[out=90,in=-135] (2.5,4);
        \node[left] at (1,2) {$b_1$};
        \draw[gray,thick] (2.5,0) to[out=45,in=-90] (4,2) to[out=90,in=-45] (2.5,4);
        \node[right] at (4,2) {$b_2$};

        \draw[blue] plot [mark=*, mark size=2] coordinates{(2.5,2.5)};
        
    \end{tikzpicture}
    \caption{Cross section through the bulk of $\tilde{\mathcal{M}}$. Branes $b_1$, $b_2$ travel in a timelike direction, meeting somewhere to the future of $E_{\mathcal{V}_1\mathcal{V}_2}$. The shaded region is a cross section of the scattering region, where the interaction takes place. Points $c_1$ and $c_2$ are out of this plane, but would be in front of and behind the cross section shown. The lower boundary of the blue shaded region is where $J^+(c_1)$, $J^+(c_2)$ meet. Outputs from the interaction pass through its two future boundaries, travelling towards $r_1$ and $r_2$ respectively. The entangling surface $\gamma_{\mathcal{W}_1}=\gamma_{\mathcal{W}_2}$ (blue dot) sits at the future boundary of $E_{\mathcal{V}_1\mathcal{V}_2}$. As such, the outputs pass into $E_{\mathcal{W}_1}$ (red) and $E_{\mathcal{W}_2}$ (orange) and do so before reaching the ETW branes.}
    \label{fig:cutaway}
\end{figure}
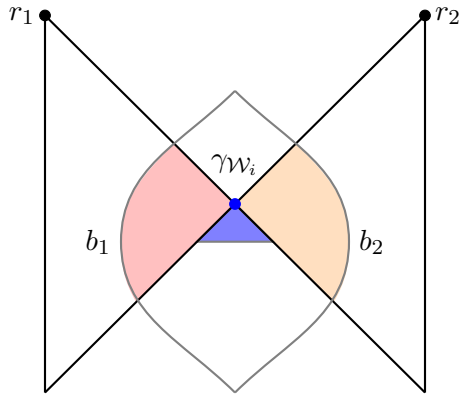

Given that the radial line $\gamma_{\mathcal{W}_1}'$ defines the wedge of $\mathcal{W}_1$ (and of its complement $\mathcal{W}_2$), there is a simple way to describe its entanglement wedge. The wedge $E_{\mathcal{W}_1}$ is just the wedge of the interval $(0,\pi)$ at time $t=\pi/2$ taken in the spacetime $\mathcal{M}$, then restricted to $\tilde{\mathcal{M}}$. 
This placement of entanglement wedges always ensures $A_1$ enters the wedge of $\mathcal{W}_1$, and $A_2$ enters the wedge of $\mathcal{W}_2$. 
To see this, observe that the two future boundaries of $E_{\mathcal{V}_1\mathcal{V}_2}$ are the past boundaries of $E_{\mathcal{W}_1}$ and $E_{\mathcal{W}_2}$. 
Thus, $A_1$, since it travels to $r_1$, enters $E_{\mathcal{W}_1}$, and similarly $A_2$ enters $E_{\mathcal{W}_2}$. 
We only need to see that this happens before these systems leave $\tilde{\mathcal{M}}$ by colliding with the branes, which again is true because the brane sits (strictly) outside of $E_{\mathcal{V}_1\mathcal{V}_2}$. 
We illustrate this in figure \ref{fig:cutaway}. 

Since $A_1$ sits in $E_{\mathcal{W}_1}$ and $A_2$ in $E_{\mathcal{W}_2}$, Alice$_1$ and Alice$_2$ can use entanglement wedge reconstruction to recover the output systems. 
Further, because we consider input systems living in a subspace small enough not to move the entangling surface, this can be done via a recovery channel which is universal for all states of the input \cite{hayden2019learning}. 

\vspace{0.2cm}
\noindent \textbf{Entanglement in the BCFT state}
\vspace{0.2cm}

So far, we have established that the state $\ket{\tilde{\Psi}}_{V_1V_2}$ suffices to perform a computation in the standard non-local scenario whenever it can be completed in the bulk scattering region. 
Next, we study how much entanglement is available in $\ket{\tilde{\Psi}}_{V_1V_2}$, and how this compares to the entanglement in the original CFT state $\ket{\Psi}_{V_1X_1V_2X_2}$. 

To relate these quantities it is helpful to recall the definition of the \emph{entanglement wedge cross section}, and its relationship to the mutual information. 
Informally, given two boundary subregions $\mathcal{A}$, $\mathcal{B}$, the entanglement wedge cross section is defined by finding the entanglement wedge of $\mathcal{A}\cup \mathcal{B}$, call it $E_{\mathcal{A}\mathcal{B}}$, and the minimal area extremal surface $\gamma$ that divides $E_{\mathcal{A}\mathcal{B}}$ into a portion homologous to $\mathcal{A}$ and portion homologous to $\mathcal{B}$. 
The entanglement wedge cross section is then
\begin{align}
    E_W(A:B) = \frac{\text{area}(\gamma)}{4G_N}.
\end{align}
In \cite{dutta2021canonical}, twice the entanglement wedge cross section was shown to be equal to an entanglement quantity known as the reflected entropy, and denoted $S_R(A:B)$, so that
\begin{align}\label{eq:SRandEW}
    S_R(A:B) = 2E_W(A:B).
\end{align}

In \cite{hayden2021markov}, it was proven that for holographic states the reflected entropy and the mutual information are related by the inequality, 
\begin{align}
    S_R(A:B) - I(A:B) \geq \frac{\log (2) \ell}{2G_N}k
\end{align}
where $k$ is the number of boundary points of $\gamma$.
In the setting we apply this we have $k=2$. 
This becomes an equality in the limit where $\mathcal{A}$, $\mathcal{B}$ occupy the entire boundary. 

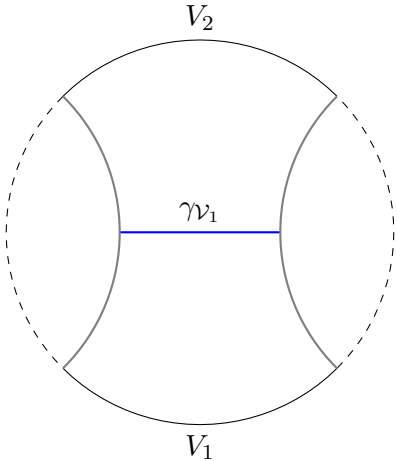
\begin{figure}
    \centering
    \begin{tikzpicture}[scale=0.85]

    \draw [domain=-45:45,black,dashed] plot ({3*cos(\x)},{3*sin(\x)});
    \draw [domain=45:135,black] plot ({3*cos(\x)},{3*sin(\x)});
    \draw [domain=135:225,black,dashed] plot ({3*cos(\x)},{3*sin(\x)});
    \draw [domain=225:315,black] plot ({3*cos(\x)},{3*sin(\x)});

    \node[above] at (0,0) {$\gamma_{\mathcal{V}_1}$};
    \draw[thick,blue] (-1.25,0) -- (1.25,0);
    
    \draw[gray, thick] (2.12, 2.12) to [out=-135,in=135] (2.12, -2.12);
    \draw[gray, thick] (-2.12, 2.12) to [out=-45,in=45] (-2.12, -2.12);
    
    \node[above] at (0,3) {$V_2$};
    \node[below] at (0,-3) {$V_1$};
        
    \end{tikzpicture}
    \caption{A constant time slice of AdS$_{2+1}$ with two ETW branes (shown in gray). The minimal surface (blue) homologous to $\mathcal{V}_1$, $\gamma_{\mathcal{V}_1}$, connects the two branes.}
    \label{fig:EWcrosssection}
\end{figure}

Consider the entanglement wedge of $\mathcal{V}_1$ in the BCFT geometry $\tilde{\mathcal{M}}$. 
The minimal surface enclosing this will be the one shown in figure \ref{fig:EWcrosssection}, that joins the two ETW branes.
Because we've considered zero tension branes, the brane trajectory corresponds to the minimal surfaces that define $E_{\mathcal{V}_1\mathcal{V}_2}$ in the original geometry $\mathcal{M}$.
The area of the minimal surface in $\tilde{\mathcal{M}}$ is then equal to the entanglement wedge cross section measured in the original geometry $\mathcal{M}$.
This leads to
\begin{align}
    I(V_1:V_2)_{\tilde{\Psi}} = 2E_W(V_1:V_2)_\Psi.
\end{align} 
This means that the mutual information in the BCFT state is equal to the canonical entropy in the original CFT state, which from equation \ref{eq:SRandEW} is twice the entanglement wedge cross section, giving 
\begin{align}
    I(V_1:V_2)_{\tilde{\Psi}} = S_R(V_1:V_2)_{\Psi}.
\end{align}
Finally, from \cite{hayden2021markov} we have that for large regions $\mathcal{V}_1$, $\mathcal{V}_2$, the canonical entropy and mutual information become close, so as $\mathcal{V}_1$ and $\mathcal{V}_2$ become large we have
\begin{align}
    I(V_1:V_2)_{\tilde{\Psi}} = S_R(V_1:V_2)_{\Psi} \rightarrow I(V_1:V_2)_{\Psi} + \frac{\log (2) \ell}{G_N}.
\end{align}
We find that the mutual information in the original CFT state and the introduced BCFT state are becoming close, up to an additive constant. 

Recapping this discussion, we saw in the last section that the entanglement in the BCFT state $\tilde{\Psi}$ suffices to reproduce in the non-local form any computations happening inside the scattering region. 
The mutual information in the BCFT state is expressed in terms of the original CFT state as the reflected entropy $S_R(V_1:V_2)_{\Psi}$. 
In the limit of large scattering regions, $S_R(V_1:V_2)_{\Psi}$ is equal to the mutual information in the original CFT state, up to an additive constant, so in that limit the mutual information in the original CFT state controls the needed entanglement to support the computation happening inside the scattering region. 

\subsection{Extension to higher dimensional global AdS}\label{sec:globalhigherdim}

In higher dimensional global AdS we can also relate bulk interactions to non-local computation, though we lose some of our quantitative statements. 
For concreteness, let's consider AdS$_{3+1}$/CFT$_{2+1}$.

\begin{figure}
    \centering
    \subfloat[\label{fig:ads2+1global}]{
    \tdplotsetmaincoords{10}{0}
    \begin{tikzpicture}[scale=1.2,tdplot_main_coords]
    \tdplotsetrotatedcoords{0}{10}{0}
    
    \begin{scope}[tdplot_rotated_coords]

    \begin{scope}[canvas is xy plane at z=0]
    \draw [smooth,domain=0:360,black] plot ({2*cos(\x)}, {2*sin(\x)});
    \end{scope}

    \begin{scope}[canvas is yz plane at x=0]
    \draw [smooth,domain=180:360,black] plot ({2*cos(\x)}, {2*sin(\x)});
    \draw [smooth,domain=0:180,black,dashed] plot ({2*cos(\x)}, {2*sin(\x)});
    \end{scope}

    \begin{scope}[canvas is xz plane at y=0]
    \draw [smooth,domain=180:360,black] plot ({2*cos(\x)}, {2*sin(\x)});
    \draw [smooth,domain=0:180,black,dashed] plot ({2*cos(\x)}, {2*sin(\x)});
    \end{scope}

    \begin{scope}[canvas is yz plane at x=2.3]
    \draw [smooth,domain=0:280,black,->] plot ({0.3*cos(-\x)}, {0.3*sin(-\x)});
    \end{scope}

    \node at (2.5,0,0) {$\theta$};

    \draw[gray,->] (0,0,0) -- (1,0,0);
    \draw[gray,->] (0,0,0) -- (0.75,0.75,0);
    \draw[->] (1,0,0) to[out=90,in=-45] (0.75,0.75,0);

    \node[above right] at (0.8,0.6) {$\phi$};

    \end{scope}
    \end{tikzpicture}}
    \hfill
    \subfloat[\label{fig:ads3+1brane}]{
    \tdplotsetmaincoords{10}{0}
    \begin{tikzpicture}[scale=1.2,tdplot_main_coords]
    \tdplotsetrotatedcoords{0}{10}{0}
    
    \begin{scope}[tdplot_rotated_coords]

    \begin{scope}[canvas is xy plane at z=0]
    \draw[smooth,domain=-75:75,black] plot ({2*cos(\x)}, {2*sin(\x)});
    \draw[smooth,domain=-75:75,black] plot ({-2*cos(\x)}, {2*sin(\x)});
    \draw[smooth,domain=10:170,black] plot ({0.5*cos(\x)}, {0.5*sin(\x)-2});
    \draw[smooth,domain=-10:-170,black] plot ({0.5*cos(\x)}, {0.5*sin(\x)+2});
    \end{scope}

    \begin{scope}[canvas is yz plane at x=0]
    \draw [smooth,domain=180:360,black] plot ({1.5*cos(\x)}, {1.5*sin(\x)});
    \draw [smooth,domain=0:180,black,dashed] plot ({1.5*cos(\x)}, {1.5*sin(\x)});
    \end{scope}

    \begin{scope}[canvas is yz plane at x=0.5]
    \draw [smooth,domain=180:360,black] plot ({1.94*cos(\x)}, {1.94*sin(\x)});
    \draw [smooth,domain=0:180,black,dashed] plot ({1.94*cos(\x)}, {1.94*sin(\x)});
    \end{scope}

    \begin{scope}[canvas is yz plane at x=-0.5]
    \draw [smooth,domain=160:410,black] plot ({1.94*cos(\x)}, {1.94*sin(\x)});
    \draw [smooth,domain=50:160,black,dashed] plot ({1.94*cos(\x)}, {1.94*sin(\x)});
    \end{scope}

    \begin{scope}[canvas is xz plane at y=0]
    \draw [smooth,domain=0:75,black,dashed] plot ({2*cos(\x)}, {2*sin(\x)});
    \draw [smooth,domain=105:180,black,dashed] plot ({2*cos(\x)}, {2*sin(\x)});
    \draw [smooth,domain=180:255,black] plot ({2*cos(\x)}, {2*sin(\x)});
    \draw [smooth,domain=285:360,black,dashed] plot ({2*cos(\x)}, {2*sin(\x)});
    \draw[smooth,domain=10:170,black] plot ({0.5*cos(\x)}, {0.5*sin(\x)-2});
    \draw[smooth,domain=-10:-170,black] plot ({0.5*cos(\x)}, {0.5*sin(\x)+2});
    
    \end{scope}
    
    \end{scope}
    \end{tikzpicture}
    }
    \caption{(a) A constant time slice of global AdS$_{2+1}$. (b) The CFT is replaced with two BCFTs defined on caps. The brane is expected to take on a connected configuration as shown when the caps are made large enough.}
\end{figure}
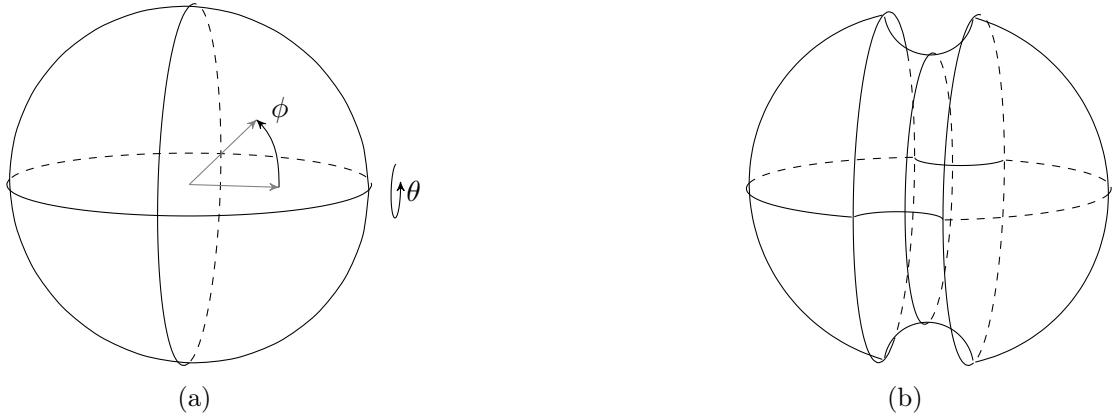

We show a single time slice of AdS$_{3+1}$ in figure \ref{fig:ads2+1global}. 
A basic obstruction to applying the ideas used above in this setting is that whenever we have in the bulk
\begin{align}
    J_{12\rightarrow 12} = J^+(c_1)\cap J^+(c_2) \cap J^-(r_1) \cap J^-(r_2) \neq \emptyset
\end{align}
we will also have $\hat{J}_{12\rightarrow 12}\neq \emptyset$. 
Thus in this setting scattering in the bulk comes along with scattering in the boundary, and we can't immediately argue entanglement in the boundary is needed. 
However, our understanding of the AdS/CFT dictionary suggests the correct picture for how the boundary reproduces the bulk interaction is not that it is implemented directly as a local interaction in the boundary scattering region. 
Instead, since the bulk scattering region is recorded only into large boundary regions, we still expect some entanglement based boundary process is supporting this bulk physics. 
We can justify this intuition using a brane construction.

To construct our brane solution, we replace the CFT with two BCFTs, defined on caps restricted to $\phi<\pi/2-\epsilon$, $\phi>\pi+\phi_0$, and look for brane solutions that end on $\phi=\pi/2\pm \epsilon$.
This corresponds to a CFT with a strip of angular width $2\epsilon$ around the equator of the sphere removed. 
As the strip becomes thin enough we expect the bulk geometry connects the two caps, leaving the bulk geometry connected, and that we see this behaviour in any number of dimensions.\footnote{This was also discussed in detail in \cite{van2020spacetime}.} 
We will assume the strip is chosen small enough such that this is the case, so that the brane solution takes on the topology shown in figure \ref{fig:ads3+1brane}. 

We can use this brane geometry to argue interactions happening in the bulk of global AdS$_{3+1}$ can be reproduced as non-local quantum computations as follows. 
We take the two caps on which our BCFTs are defined, call them $\mathcal{C}_L$ and $\mathcal{C}_R$, to be our input regions, and then define output regions
\begin{align}
    \mathcal{R}_+ &= \{q:0<\theta<\pi, t=T \} \nonumber \\
    \mathcal{R}_- &= \{q:-\pi<\theta<0, t=T \}
\end{align}
to be the output regions. 
We define the scattering region as before, 
\begin{align}
    J_{LR\rightarrow +-} = J^+(\mathcal{C}_R) \cap J^+(\mathcal{C}_L) \cap J^-(\mathcal{R}_+)\cap J^+(\mathcal{R}_-)
\end{align}
which captures the portion of the bulk where interactions can be reproduced in the non-local form. 
Explicitly, to do this one inserts input systems into the bulk by acting on the input regions, and has them interact and send outputs towards the $+$ and $-$ regions.
Using the communication round of the non-local computation to exchange degrees of freedom, the second round operations then act on the $+$ and $-$ regions to recover the outputs. 

It remains to argue that, at least for $\epsilon$, $T$ small enough, the scattering region is non-empty. 
To see this, first notice that if we make $\epsilon$, $T$ small enough the extremal surface attached to $\theta=0$ and $\theta=\pi$, $t=T$ will connect the two BCFTs. 
We claim that whenever this is the case at least the point $p$ at $t=r=0$ is inside the scattering region. 
To see this, note that the extremal surface enclosing the left (right) BCFT at $t=0$ is the $\phi=\pi/2$ plane, which includes $p$, so
\begin{align}
    p \in J^+(\mathcal{C}_L) \cap J^+(\mathcal{C}_R).
\end{align}
It remains to show $p$ is inside the past of the wedges $E_{\mathcal{R}_-}$, $E_{\mathcal{R}_+}$.
To see this, define the regions 
\begin{align}
    \mathcal{R}_+' &= \{q:0<\theta<\pi, t=-T \} \nonumber \\
    \mathcal{R}_-' &= \{q:-\pi<\theta<0, t=-T \}
\end{align}
Notice that the entangling surface for $E_{\mathcal{R}_+}$ must sit at $t>0$ when it reaches $r=0$. 
If it didn't, by symmetry it would cross the entangling surface for $E_{\mathcal{R}'_+}$, which can't happen by causality. 
Similarly, the entangling surface for $E_{\mathcal{R}_-}$ must sit at $t>0$ when it reaches $r=0$. 
But then both entangling surface have $p$ in their past, so 
\begin{align}
    p \in J^+(\mathcal{C}_L) \cap J^+(\mathcal{C}_R) \cap J^-(\mathcal{R}_-) \cap J^-(\mathcal{R}_+)
\end{align}
and hence the scattering region is non-empty. 

\section{Discussion}\label{sec:discussion}

A lesson of \cite{may2019quantum,may2020holographic,may2022connected} is that boundary entanglement plays a necessary role in supporting bulk interaction. 
Non-local computation provides a quantum information theoretic way to understand this, and in particular to explore the many questions around what interactions can be supported by entanglement in this way, and conversely what holography implies about non-local computation. 
In this work, we extended this basic lesson to higher dimensions and two sided black holes. 

With briefly conclude with a few comments on future directions or relationships to other work.  

\vspace{0.2cm}
\noindent \textbf{Complexity and the black hole singularity}
\vspace{0.2cm}

There is a general expectation that high complexity unitaries should require large entanglement to implement in the non-local form. 
Assuming this, our construction would give that high complexity unitaries are forbidden from being implemented inside reasonably small sized subregions of the black hole. 
From a bulk perspective, this is a unsurprising claim: we might expect high complexity operations to require large physical time to implement,\footnote{For another perspective on this, see \cite{kubicki2023constraints}.} and there is finite time before reaching the singularity.

What is perhaps more interesting is that the non-local computation perspective suggests a boundary understanding of the appearance of this finite time, and hence on the appearance of this singularity. 
We suggest that the CFT is limited by its finite entanglement to only allow low complexity computations to happen, and the bulk dual ``geometrizes'' this constraint via the appearance of the bulk singularity, which ends bulk time and limits computation.\footnote{We thank Steve Shenker for comments made to us in this direction.} 
Unfortunately, our techniques so far fail to capture interactions happening in the deep interior, in particular interactions happening past the last extremal surface. 
Consequently, entanglement based constraints on non-local computation would only constrain interactions happening away from the singularity.\footnote{See however \cite{kubicki2023constraints} for some more primitive constraints that do apply to the deep interior region.} 
In future work, we hope to better understand interactions in the deep interior and relate this to the appearance of the singularity. 

\vspace{0.2cm}
\noindent \textbf{Efficiency of non-local computation and holography}
\vspace{0.2cm}

An interesting tension between the efficiency of the current best non-local computation strategies and expectations of what it should be possible to compute in a subregion of the bulk was raised in \cite{may2020holographic} and explored in \cite{may2022complexity}. 
One way out of this tension was, previously, to recall that holography is most naturally related to `augmented' non-local computations, and that entanglement between the input regions is actually unnecessary in these augmented non-local computations. 
While it was argued earlier that the standard non-local computation scenario is the relevant one on heuristic grounds in \cite{may2020holographic,may2022complexity}, and by going to the lattice setting in \cite{dolev2022holography}, our construction gives another sharp setting where the efficiency of existing non-local computation protocols is in tension with having reasonable bulk computations. 
Another route out of this tension was previously to point to the connection between non-local computation and holography being most precise in low dimensions, where it would be less surprising if bulk interactions were severely limited. 
This route out of the tension is now also closed. 

Because non-local quantum computations are cheating strategies in the cryptographic setting of position-verification \cite{buhrman2014position}, the suggestion from holography that many computations should be efficiently implementable has potential practical implications. 
Conversely, an interesting possibility is to use constraints on non-local computation to constrain bulk physics, and in particular constrain the complexity of computation happening in the bulk \cite{may2022complexity}. 
Our construction further supports that such entanglement constraints should constrain bulk physics. 

\vspace{0.2cm}
\noindent \textbf{Non-local computation with fixed second round}
\vspace{0.2cm}

As noted in section \ref{sec:generaldblackholecase}, we related bulk interactions to non-local computations where the second round is fixed, in the sense that the second round doesn't depend on the choice of interaction or on the inputs. 
Non-local computation with a fixed second round has already appeared in the quantum information literature. 
In \cite{junge2022geometry} the authors assumed a fixed second round and derived interesting new constraints on non-local computation, under some mathematical conjectures on type constants in Banach spaces. 
Fixed second rounds also appear naturally elsewhere in cryptography \cite{kawachi2021recent}\footnote{In this reference the analogous idea is of ``universal reconstruction''.}. 
We were not able to connect the constraints of \cite{junge2022geometry} directly to holography and our setting, but this work does suggest that adding that the second round is fixed may allow for stronger constraints on bulk computation to be understood. 
Another comment is that all non-local computation protocols devised so far do involve fixed second rounds, and the observation here is just that protocols built from studying the holographic setting also share this property. 

\vspace{0.2cm}
\noindent \textbf{Sub-AdS scale scattering regions}
\vspace{0.2cm}

Examining the connection between non-local computation and bulk interactions from the perspective of black holes highlights an issue relating to sub-AdS locality. 
In both the two-sided and one-sided cases, our boundary description of the bulk interaction includes a description of a super-AdS sized subregion of the bulk, even when the interaction is localized to a small region. 
In our context, this is enforced by a change in the minimal extremal surface homologous to certain boundary subregions. 
In this sense, our quantum  circuit model of behind the horizon interactions can't (at this stage) isolate the boundary description of interactions happening within sub-AdS scale bulk regions. 
The construction in \cite{dolev2022holography} also suffers from this limitation. 
This is analogous to tensor network models of spatial slices of the bulk geometry, which also struggle to capture sub-AdS locality \cite{bao2015consistency,pastawski2015holographic,hayden2016holographic,yang2016bidirectional}.
However, the heuristic arguments of \cite{may2019quantum,may2021holographic,may2022connected} suggest that there should be a non-local computation description of sub-AdS scale bulk subregions.\footnote{In particular, the connected wedge theorem of \cite{may2019quantum,may2021holographic,may2022connected} gives that the boundary entanglement is consistent with small bulk subregions having a non-local computation description.} 
This motivates looking for constructions that can related the sub-AdS sized regions to (non-augmented) non-local computation. 
We comment in the discussion on some extensions of our setting that may allow this.  

To construct sub-AdS scale bulk regions whose interactions are supported as a non-local computation within our setting, we would need to find geometries where our branes can come within an AdS distance of each other while remaining the dominant saddle in the path integral, and while keeping the connected (non-brane anchored) extremal surface minimal. 
Towards keeping the connected brane geometry dominant, we could consider variants of our geometries involving charged black hole and charged branes, and we could consider branes with non-zero (and perhaps differing) tensions, and we could adjust the black hole temperature, which here was implicitly fixed. 
This would give a richer phase space for the phase transition from connected to disconnected branes, potentially allowing the branes to come within a sub-AdS distance of each other while remaining disconnected. 
To keep the connected extremal surface minimal, we could consider adding a dilaton field to the branes, which can be used to raise the generalized entropy associated with minimal surfaces ending on the brane \cite{almheiri2020page}.  

\vspace{0.2cm}
\noindent \textbf{ETW brane geometries and information processing}
\vspace{0.2cm}

The work \cite{van2020spacetime} proposes entangled BCFTs as an alternative set of degrees of freedom that, prepared in appropriate states, can approximate holographic CFTs. 
An advantage to this alternative description is the finite entanglement among subsystems. 
Combined with the perspective of \cite{bao2019beyond} on compressing holographic states, this allowed us to move from a description of a quantum information processing protocol in a holographic CFT, involving infinite dimensional systems, to a (approximate) description using only finite dimensional systems. 

Finding a finite dimensional description of these holographic protocols was discussed in \cite{dolev2022holography} under an assumption the holographic CFT is well approximated by a lattice system, and more commonly relating AdS/CFT to quantum information processing protocols has been discussed in the context of tensor network toy models. 
AdS/BCFT combined with compression results has a set of advantageous features not reproduced in those settings: it gives a concrete model for what bulk physics is being captured (the original geometry, ended by an ETW brane), can be justified and studied directly from the standpoint of the Euclidean path integral, and has top-down realizations. 
Moving beyond non-local computation, this seems a promising approach to relating other quantum information processing protocols to holography.
Another recent example of AdS/BCFT being used to model interesting information theoretic manipulations of holographic CFTs is given in \cite{antonini2022holographic,antonini2023holographic}. 

\vspace{0.2cm}
\noindent \textbf{Acknowledgements:} We thank Patrick Hayden, David P\'erez-Garcia, Shreya Vardhan, Henry Lin, Stefano Antonini, and Raghu Mahajan for helpful discussions. AM is supported by the Simons Foundation It from Qubit collaboration, a PDF fellowship provided by Canada's National Science and Engineering Research council, and by Q-FARM. MX is supported by an NSF Graduate Research Fellowship and the Simons Foundation.

\appendix 

\section{Coordinate systems}\label{sec:coordinatechanges}

We summarize the coordinate systems used in this article. 
We describe our coordinates and their relationships using the embedding space formalism. 
In particular, our coordinates are parameterizations of the surface $X^AX_A=X_0^2-X_1^2 - X_2^2 + X_3^2=\ell^2$, in the metric $\text{diag}(1,-1,-1,1)$. 

Lorentzian Poincare AdS$_{2+1}$ is described by coordinates.
\begin{align}
    X_0 &= \frac{z}{2}\left(1 + \frac{\ell^2+x^2-t^2}{z^2} \right) \nonumber \\
    X_1 &= \frac{z}{2}\left(1 - \frac{\ell^2-x^2+t^2}{z^2} \right) \nonumber \\
    X_2 &= \ell\frac{x}{z} \nonumber \\
    X_3 &= \ell\frac{t}{z} .
\end{align}

Lorentzian global AdS$_{2+1}$ is described by
\begin{align}
    X_0 &= \ell \frac{\cos(t)}{\cos(r)} \nonumber \\
    X_1 &= \ell \tan(r) \sin(\phi) \nonumber \\
    X_2 &= \ell \tan(r) \cos(\phi) \nonumber \\
    X_3 &= \ell \frac{\sin(t)}{\cos(r)}.
\end{align}

The Kruskal coordinates of the Lorentzian BTZ black hole are described by
\begin{align}
    X_0 &= \ell \frac{1-uv}{1+uv}\cosh\frac{R\phi}{\ell} \nonumber\\ 
    X_1 &= \ell \frac{1-uv}{1+uv}\sinh\frac{R\phi}{\ell} \nonumber\\ 
    X_2 &= \ell \frac{v-u}{1+uv} \nonumber \\ 
    X_3 &= \ell \frac{v+u}{1+uv} 
\end{align}

The global, Lorentzian coordinates of the BTZ black hole are
\begin{align}
    X_0 &= \ell \cos(s) \sec(w) \cosh(X)\nonumber \\
    X_1 &= \ell \cos(s) \sec(w) \sinh(X) \nonumber \\
    X_2 &= \ell \tan(w) \nonumber \\
    X_3 &= \ell \sin(s) \sec(w).
\end{align}
These describe the standard BTZ black hole when $X$ is periodic, and planar BTZ when $X$ is an interval.

One exterior region of the planar BTZ black hole is covered by the coordinates
\begin{align}
    X_0 &= \ell \sqrt{1+\frac{\rho^2}{\ell^2}}\cosh X \nonumber \\
    X_1 &= \ell \sqrt{1+\frac{\rho^2}{\ell^2}}\sinh X \nonumber \\
    X_2 &= \rho \cosh t \nonumber \\
    X_3 &= \rho \sinh t .
\end{align}

\section{Brane in Poincare AdS\texorpdfstring{$_{2+1}$}{TEXT}}\label{appendix:Poincare2+1}

\subsection{Single brane solution}

\begin{figure}
    \centering
    \subfloat[\label{fig:puncturedplane}]{
    \tdplotsetmaincoords{15}{0}
    \begin{tikzpicture}[scale=1,tdplot_main_coords]
    \tdplotsetrotatedcoords{0}{-30}{0}
    \begin{scope}[tdplot_rotated_coords]

    \begin{scope}[canvas is xy plane at z=0]
    \draw[thick,black,fill=blue,opacity=0.3] (-3,-3) -- (-3,3) -- (3,3) -- (3,-3) -- cycle;
    \draw[thick,black,fill=white] (0,0) circle[radius=1.5] ;
    \end{scope}

    \end{scope}
    \end{tikzpicture}}
    \hfill
    \subfloat[\label{fig:Poincarebrane}]{
    \tdplotsetmaincoords{15}{0}
    \begin{tikzpicture}[scale=1,tdplot_main_coords]
    \tdplotsetrotatedcoords{0}{-30}{0}
    
    \begin{scope}[tdplot_rotated_coords]

    \begin{scope}[canvas is xy plane at z=0]
    \draw [domain=270:360,black,dashed] plot ({1.5*cos(\x)},{1.5*sin(\x)});
    \draw [domain=0:65,black,dashed] plot ({1.5*cos(\x)},{1.5*sin(\x)});
    \draw[thick,black,fill=blue,opacity=0.3] (-3,-3) -- (-3,3) -- (3,3) -- (3,-3) -- cycle;
    \fill[white] (0,0) circle[radius=1.5] ;
    \draw [domain=65:270,black] plot ({1.5*cos(\x)},{1.5*sin(\x)});
    
    \end{scope}

    \draw [domain=180:360,black] plot ({1.5*cos(\x)}, 0,{1.5*sin(\x)});
    \draw [domain=180:360,black] plot (0,{1.5*cos(\x)},{1.5*sin(\x)});

    \foreach \i in {0,...,40}
    {
    \begin{scope}[canvas is xy plane at z={-1.5*\i/40}]
    \draw[black,opacity=0.2] (0,0) circle[radius={(1.5^2-(1.5*\i/40)^2)^0.5}] ;;
    \end{scope}
    }
    
    \end{scope}
    \end{tikzpicture}
    }
    \caption{(a) The Euclidean CFT path integral with boundary conditions set along a disk at $x^2+\tau^2=R^2$. (b) The dual bulk geometry, with a zero tension brane at $x^2+\tau^2 + z^2 = R^2$.}
\end{figure}
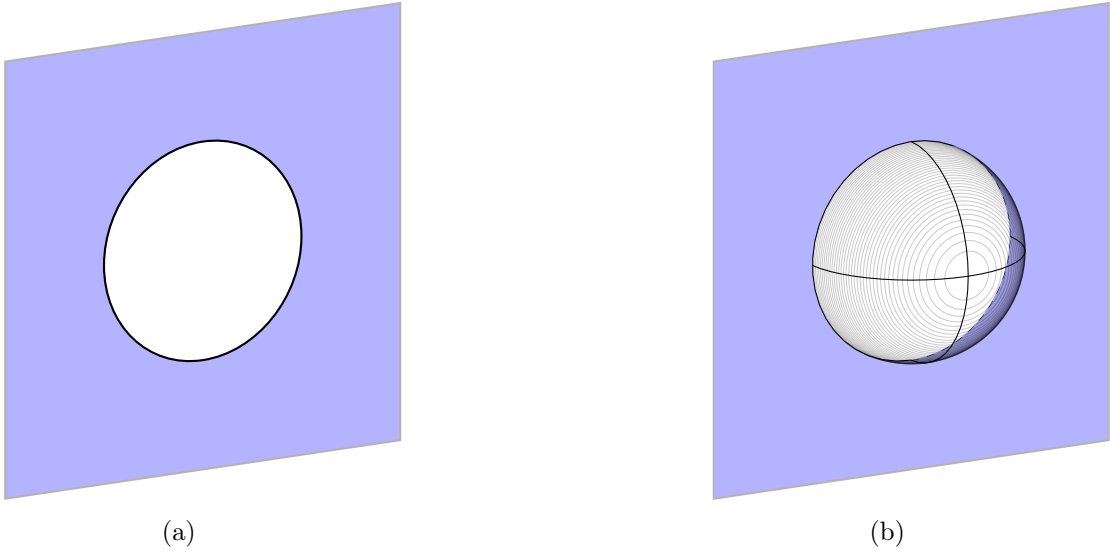

The simplest brane solution we study is dual to the Euclidean path integral shown in figure \ref{fig:puncturedplane}. 
There, we are considering the Euclidean CFT path integral on the plane with no operator insertions, and a boundary condition set at $x^2+\tau^2=R^2$. 
We will choose this boundary condition such that the boundary entropy is zero. 
The bulk dual then will be a zero tension brane extending into the bulk and meeting the asymptotic boundary at $x^2+\tau^2=R^2$. 

The solution was studied already in \cite{rozali2020information}, who found a Euclidean Poincare AdS$_{3}$ bulk metric
\begin{align}
    ds^2 = \frac{\ell^2}{z^2}(dx^2 + d\tau^2 + dz^2)
\end{align}
and brane trajectory defined by
\begin{align}
    x^2 + z^2 + \tau^2 = R^2.
\end{align}
Variations of this solution, obtained by coordinate transformations, adding a second circular boundary, or Wick rotating appear throughout this article. 
One Lorentzian brane solution we will use is
\begin{align}
    x^2 +z^2 - t^2 = R^2.
\end{align}
With the metric being Lorentzian Poincare AdS$_{2+1}$.
This is obtained by Wick rotating the $\tau$ coordinate. 

\subsection{Minimal surfaces}

Note that for $R=1$ these results can be extracted from \cite{rozali2020information}. 
We are just modifying their calculation to allow arbitrary $R$, and showing the calculation directly in Lorentzian signature.

\begin{figure}
    \centering
    \begin{tikzpicture}[scale=1]

    \draw[gray,->] (0,0) -- ({2*cos(120)}, {2*sin(120)});
    \node[left] at (-0.5,0.65) {$R$};
    
    \draw[thick] (-3,0) -- (10,0);
    \draw [domain=0:180,gray,thick] plot ({2*cos(\x)}, {2*sin(\x)});
    \draw[black] plot [mark=*, mark size=2] coordinates{(0,0)};
    \node[below] at (0,0) {$x=0$};
    
    \draw[black] plot [mark=*, mark size=2] coordinates{(7,0)};
    \node[below] at (7,0) {$x=L$};
    
    \draw [domain=0:148,blue] plot ({3.21*cos(\x)+7-3.21}, {3.21*sin(\x)});
    
    \draw [domain=180:148,gray] plot ({0.65*cos(\x)+7-3.21}, {0.65*sin(\x)});
    \node at (2.9,0.25) {$\alpha$};
    
    \draw[gray,->] ({7-3.21},0) -- ({3.21*cos(60)+7-3.21}, {3.21*sin(60)});
    \node at (5,1.45) {$r_H$};

    \draw[gray] ({7-3.21},0) -- ({3.21*cos(148)+7-3.21}, {3.21*sin(148)});
    \draw[black] plot [mark=*, mark size=2] coordinates{({7-3.21},0)};
    
    \end{tikzpicture}
    \caption{The $t=0$ slice of Euclidean Poincare AdS$_3$. An ETW brane (gray) centered at $x=0$ sits at $x^2+z^2=R^2$. A minimal surface (blue) anchored to $x=L$ is a portion of a semi-circle and meets the brane orthogonally.}
    \label{fig:geometryfig}
\end{figure}
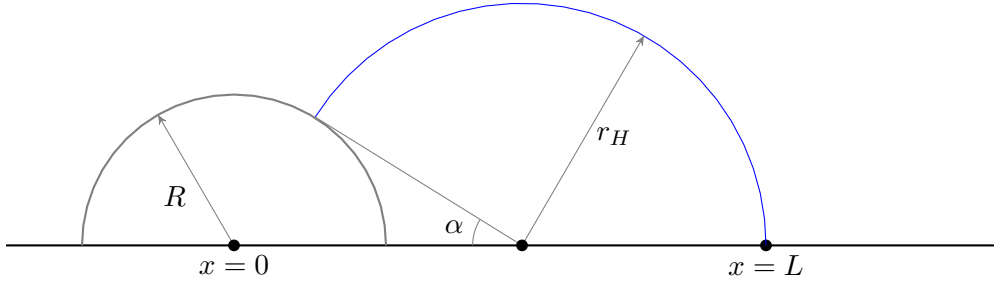

We start with minimal surfaces in the $t=0$ slice of Lorentzian Poincare AdS. 
The brane is sitting at
\begin{align}
    x^2 + z^2 = R^2.
\end{align}
Consider an interval extending from the brane to $x=L$. 
The minimal surface extending from this point is a portion of a circle, meeting the brane orthogonally. 
In figure \ref{fig:geometryfig}, we do some elementary geometry to determine the radius of this circle and the angle it extends over. 

We find
\begin{align}\label{eq:meetingparameters}
    r_H &= \frac{L^2-R^2}{2L}, \nonumber \\
    \tan \alpha &= \frac{2L R}{L^2-R^2}.
\end{align}
Thus the geodesic is given parametrically by
\begin{align}
    x(\theta) &= L - r_H + r_H \cos(\theta), \nonumber \\
    t(\theta) &= 0, \nonumber \\
    z(\theta) &= r_H \sin(\theta).
\end{align}
where $0\leq \theta \leq \pi-\alpha$.

Calculating the area of these curves, we find, for the area of a geodesic starting at $x=x_0,t=0$,
\begin{align}
    A[x_0,0] = \ell \log \left( \frac{x_0^2 - R^2}{R\delta }\right)
\end{align}
Note that here we use a cut-off at $z=\delta$, which means $\theta = \delta / r_H$. 

Observing that the brane trajectory is invariant under boosts in the $(x,t)$ plane, we can obtain the area of geodesics starting at general coordinates $x_0,t_0$ by writing the area in terms of the invariant interval, 
\begin{align}\label{eq:poincareareaformula}
    A[x_0,t_0] = \ell \log \left( \frac{x_0^2 -t_0^2 - R^2}{R\delta }\right)
\end{align}
We will make use of this below, by exploiting a local equivalence of the ETW branes in the planar BTZ black hole to this Poincare AdS brane. 

\section{BTZ black hole}\label{appendix:periodic_BTZ}

\subsection{Geometry}

Consider a Euclidean path integral on a torus, with geometry
\begin{align}
ds^2 = dX^2 + d\phi^2, \qquad \phi \in[0,\beta), \qquad X\in[-\pi,\pi).
\end{align}
As noted in \cite{maloney2010quantum, maldacena1999ads3}, there is a family of bulk saddles that correspond to this boundary geometry, with a distinct saddle for each matrix in $\text{SL}(2,\mathbb{Z})$.
At large $1/G_N$ only thermal AdS or the BTZ black hole appear, with the black hole solution dominating for
\begin{align}
\beta <2\pi.
\end{align}

\vspace{0.2cm}
\noindent \textbf{Coordinate change}
\vspace{0.2cm}

After Wick rotation, the Lorentzian BTZ black hole can be represented in Kruskal coordinates, which we choose because it is straightforward to relate to our preferred metric. 
The Kruskal coordinates metric is
\begin{align}
    ds^2 = \frac{-4\ell^2 dudv+R^2(1-uv)^2d\phi^2}{(1+uv)^2}.
\end{align}
with $R=2\pi \ell/\beta$.
We would like to turn this into the following metric for our calculations to follow:
\begin{align}\label{eq:appendixglobalBHmetric} 
ds^2 = \frac{\ell^2}{\cos^2(w)}\left(-ds^2+dw^2+\cos^2(s)dX^2\right).
\end{align}
with $s,w\in[-\pi/2,\pi/2]$. 

We can relate the two metrics via the embedding coordinates given in section \ref{sec:coordinatechanges}, and we find
\begin{align}
    X_0/\ell &= \frac{1-uv}{1+uv}\cosh\frac{R\phi}{\ell} = \cos(s) \sec(w) \cosh(X)\nonumber \\
    X_1/\ell &= \frac{1-uv}{1+uv}\sinh\frac{R\phi}{\ell}  = \cos(s) \sec(w) \sinh(X)\nonumber \\
    X_2/\ell &= \frac{v-u}{1+uv} =  \tan(w) \nonumber \\
    X_3/\ell &= \frac{v+u}{1+uv} = \sin(s) \sec(w).
\end{align}
Notice that $R\phi/\ell=X$, so that periodicity in $\phi$ implies $X\in (-R\pi/\ell,R\pi\ell)$. 
We define $X^0=R \pi/\ell = 2\pi^2/\beta$. 

\subsection{Minimal surfaces}

Once we are working with these coordinates, we would like to understand the extremal surface candidates anchored to $w=\pm \pi/2$, $s=T$, $X=0$ and $w=\pm \pi/2$, $s=T$, $X=X^0$, and when each is minimal.

The first possible extremal surface has two pieces, which each go through the black hole and connect the two boundaries. One connects $w=\pm \pi/2$, $s=T$, $X=0$ and the other, $w=\pm \pi/2$, $s=T$, $X=X^0$. This extremal surface is the configuration we need to perform reconstruction for our protocol, since it divides our torus into a "left" and "right". In global coordinates, the minimal surface attached to $w=\pm \pi/2$, $s=T$, for constant $X$ is
\begin{align}
    s &= T, \nonumber \\
    w &= \arcsin(\tanh(\lambda)).\label{eq:connectedgeodesic}
\end{align}
The area, setting a cut-off $w\in[-\pi/2+\epsilon,\pi/2-\epsilon]$, for one piece is
\begin{align}\label{eq:connectedBTZarea}
    A_{c1}[T] &=  2 \ell \log\left(\frac{2}{\epsilon} \right).
\end{align}
Hence the total area of this extremal surface is twice that:
\begin{align}
    A_{c}[T] &=  4 \ell \log\left(\frac{2}{\epsilon} \right).
\end{align}

The second extremal surface candidate has two pieces, each of which ends back on the boundary it starts on. One connects $w=\pi/2$, $s=T$, $X=0$ with $w=\pi/2$, $s=T$, $X=X^0$, and the other connects $w=-\pi/2$, $s=T$, $X=0$ to $w=-\pi/2$, $s=T$, $X=0$. To find the area of this extremal surface, we will first find the solution in Poincare coordinates, and then translate it back into the global BTZ coordinates. From the coordinate transformations in section \ref{sec:coordinatechanges}, the global BTZ coordinates are related to Lorentzian Poincare coordinates by the transformation
\begin{align}
    t &= \ell \,e^X \tan(s), \nonumber \\
    x &= \ell \,e^X \sec(s) \sin(w), \nonumber \\
    z &= \ell \,e^X \sec(s) \cos(w) .\label{eq:globalbtztopoincare}
\end{align}
The two pieces have the same area due to symmetry, so we will just find the area of one piece and double it. The endpoints of one geodesic in Poincare coordinates are
\begin{align}
t_1 = \ell \tan T &\qquad t_2 =\ell e^{X^0} \tan T\nonumber\\
x_1 = \ell \sec T &\qquad x_2 =\ell e^{X^0} \sec T
\end{align}
The locations of these endpoints can vary wildly depending on the value of $X^0$ we pick, so the problem is essentially that of finding the area of the geodesic connecting two arbitrary points $(x_1, t_1)$ and $(x_2, t_2)$ in the Poincare metric. Notice that we can always shift the origins of $t$ and $x$ such that
\begin{align}
\frac{x_1+x_2}{2}=0 \qquad \frac{t_1+t_2}{2}=0.
\end{align}
Then we can boost the two points to the $t=0$ slice, where we know the geodesic is a semicircle and can be written as
\begin{align}
R_0^2 &= x^2+z^2, \qquad \text{for }2R_0=\sqrt{(x_2-x_1)^2-(t_2-t_1)^2}
\end{align}
Note that the parameterization of this geodesic represented in the $t=0$ slice is not necessarily affine, and that the trajectory does change under these transformations. However, we are only looking for the area, and that remains the same. For a cutoff $z=\delta$, the area from the cutoff up to the top of the semicircle is
\begin{align}
A_{\delta}=\ell \log\left(\frac{2R}{\delta}\right).
\end{align}
Keeping the BTZ cut-off at $w = \pi/2-\epsilon$, the corresponding $z$ cutoffs are $z_1 = \ell \epsilon\sec T$ for one endpoint and $z_2=\ell\epsilon e^{X^0}\sec T$ for the other. We then find the total area of our Poincare geodesic to be
\begin{align}
A_{\delta_1+\delta_2}[T] &= \ell \log\left(\frac{2R}{z_1}\right)+\ell \log\left(\frac{2R}{z_2}\right) \nonumber\\
&= 2\ell \log\left(\frac{2\sinh \left(\frac{|X^0|}{2}\right)\cos(T)}{\epsilon} \right).
\end{align}
The BTZ extremal surface has two such pieces, so the final area for this extremal surface candidate is
\begin{align}
A_d[T] = 4\ell \log\left(\frac{2\sinh \left(\frac{|X^0|}{2}\right)\cos(T)}{\epsilon} \right).
\end{align}

To be in the phase where the extremal surface connects the two boundaries, then, we need
\begin{align}
    4\ell \log\left(\frac{2\sinh\frac{|X^0|}{2} \cos T}{\epsilon} \right) -4\ell \log\left(\frac{2}{\epsilon} \right) \geq 0
\end{align}
which simplifies to
\begin{align}
    \boxed{\sinh\frac{|X^0|}{2}\geq \sec T}.
\end{align}

\section{Planar BTZ black hole}\label{appendix:BTZ}

\subsection{Brane trajectories}

Consider the Euclidean path integral on a finite cylinder,
\begin{align}
    ds^2 = dX^2 + d\phi^2 \qquad \phi \in[0,2\pi)
\end{align}
with boundary conditions set at $X=\pm X^0$. 
Choose the boundary conditions to have zero boundary entropy, dual to a bulk solution with a zero tension brane. 

There are two possible bulk solutions corresponding to this Euclidean path integral, which correspond to having either two separate branes ending on each CFT edge, or a single brane that attaches the two edges together. 
Putting boundary conditions at $X=\pm X^0$, a result of \cite{cooper2019black} (see equation A.24 there) implies the disconnected solution is the minimal action one for 
\begin{align}\label{eq:X0conditionforaction}
    X^0 \geq \frac{\pi}{2}.
\end{align}
With this condition, the bulk metric dual to the path integral on the cylinder is
\begin{align}\label{eq:appendixEuclideanBH}
    ds^2 = \ell^2\left(\frac{\rho^2}{\ell^2}+1\right)dX^2 + \frac{d\rho^2}{\frac{\rho^2}{\ell^2}+1} + \rho^2 d\phi^2,
\end{align}
and the brane trajectory is just
\begin{align}
    X = \pm X^0.
\end{align}

\vspace{0.2cm}
\noindent \textbf{Lorentzian solution}
\vspace{0.2cm}

Assuming we are in the disconnected phase, Wick rotate $\phi\rightarrow it$ in the metric \ref{eq:appendixEuclideanBH} to obtain
\begin{align}
    ds^2 = \ell^2\left(\frac{\rho^2}{\ell^2}+1\right)dX^2 + \frac{d\rho^2}{\frac{\rho^2}{\ell^2}+1} - \rho^2 dt^2
\end{align}
This covers one exterior region of the black hole. 
A global metric for this black hole, covering the full spacetime, is
\begin{align}
    ds^2 = \frac{\ell^2}{\cos^2(w)}\left(-ds^2 + dw^2 + \cos^2(s)dX^2 \right)
\end{align}
where $s\in[-\pi/2,\pi/2]$, $w\in [-\pi/2,\pi/2]$. Notice that this is the same metric as \ref{eq:appendixglobalBHmetric}, except $X$ is not periodic.
From section \ref{sec:coordinatechanges} we have that these two coordinate systems are related by
\begin{align}
    X_0/\ell &= \sqrt{1+\frac{\rho^2}{\ell^2}}\cosh X = \cos(s) \sec(w) \cosh(X)\nonumber \\
    X_1/\ell &= \sqrt{1+\frac{\rho^2}{\ell^2}}\sinh X = \cos(s) \sec(w) \sinh(X)\nonumber \\
    X_2/\ell &= \frac{\rho}{\ell} \cosh t =  \tan(w) \nonumber \\
    X_3/\ell &= \frac{\rho}{\ell} \sinh t = \sin(s) \sec(w)
\end{align}
Note that in particular the $X$ coordinates in the two spacetimes are identified trivially. 

\subsection{Minimal surfaces}

We'll use the global coordinates for the two sided planar BTZ black hole, 
\begin{align}
    ds^2 = \frac{\ell^2}{\cos^2(w)}\left(-ds^2 + dw^2 + \cos^2(s)dX^2 \right).
\end{align}
Recall these are related to Lorentzian Poincare coordinates by the transformation in equation \ref{eq:globalbtztopoincare}.

We would like to understand when the connected surface (the one that threads through the black hole) is minimal. For endpoints $w=\pm \pi/2$, $s=T$ at constant $X=0$, the geodesic takes the same form as just one piece of the periodic BTZ extremal surface that connects the two boundaries, which can be found in equation \ref{eq:connectedgeodesic}. The area is similarly just that of one piece of the periodic case, which is equation \ref{eq:connectedBTZarea}. This case shows the usefulness of using an almost identical metric for both the periodic and planar case. 

Next, we look for the trajectory of the brane anchored geodesic. 
We will again approach this by finding the solution in Poincare, and transforming to global coordinates. 
In Poincare, the endpoint of our geodesic is
\begin{align}
    x_0 &= \pm l\sec(T) \nonumber \\
    t_0 &= l\tan(T)
\end{align}
and the brane radius $R$ is related to the position of the brane in the BTZ black hole geometry by
\begin{align}
    X^0 = \ln \frac{R}{\ell}.
\end{align}
We would like to use the area formula \ref{eq:poincareareaformula} to determine the area of these brane attached surfaces. The $w=\pi/2-\epsilon$ cutoff translates to 
\begin{align}
    z=\delta = l\epsilon \sec(T).
\end{align}
Now using the area formula for the disconnected geodesic, we have
\begin{align}
    A_d[T] = 2\ell\log\left(\frac{2\sinh |X^0|\cos(T)}{\epsilon} \right).
\end{align}
To be in the connected phase, we impose
\begin{align}
    2 \ell \log\left( \frac{2\sinh |X^0| \cos T}{\epsilon} \right) -2\ell \log\left(\frac{2}{\epsilon} \right) \geq 0
\end{align}
which simplifies to
\begin{align}
    \boxed{\sinh|X^0|\geq \sec T}.
\end{align}

\section{Global AdS\texorpdfstring{$_{2+1}$}{TEXT}}\label{appendix:global2+1}

\subsection{Euclidean brane trajectories}

We are interested in finding branes ending the Euclidean global AdS spacetime
\begin{align}
    ds^2 = \frac{\ell^2}{\cos^2(r)}(d\tau_G^2 + dr^2 + \sin^2(r) d\phi^2).
\end{align}
Heuristically, we would like solutions that have boundary conditions set along roughly circular curves located antipodally on the cylinder. 
To construct a precise brane geometry, consider the coordinate transformation to Poincare AdS, given by
\begin{align}\label{eq:globaltoPoincareEuclidean}
    \tan(r) &= \frac{\sqrt{x^2+\tau^2}}{z}, \nonumber \\
    \tanh(\tau_G) &= \frac{z^2 + x^2+\tau^2 -\ell^2 }{z^2+x^2+\tau^2+\ell^2}, \nonumber \\
    \tan(\phi) &= \frac{\tau}{x}.
\end{align}
Or, inverting this,
\begin{align}
    \tau &= \ell \sin(r)\sin(\phi)e^{\tau_G}, \nonumber \\
    x &= \ell \sin(r)\cos(\phi)e^{\tau_G}, \nonumber \\
    z &= \ell \cos(r) e^{\tau_G}.
\end{align}
The resulting Poincare AdS metric is
\begin{align}
    ds^2 = \frac{\ell^2}{z^2}(dz^2 + dx^2+d\tau^2).
\end{align}

We will choose a very simple placement of the boundary conditions in Poincare AdS, which will map to branes in global AdS with appropriate qualitative features, and in particular will give the brane trajectory stated in the main text as equation \ref{eq:branetrajectory}.
In Poincare, we set boundary conditions at
\begin{align}
    (x \pm \ell x_0)^2 + \tau^2 = R^2
\end{align}
There is a solution consisting of two separate hemispheres attached to each of these edges, and a solution where the brane connects the two edges. 
We would like to understand when the disconnected solution has minimal action.
By doing a conformal transformation, we can observe that this phase transition is the same one as was studied in \cite{cooper2019black}, and which we also exploited in appendix \ref{appendix:BTZ}. 
In particular we can map the exterior of the two disks to the finite cylinder. 
The disk radii and separation fix the cylinder height. 

\vspace{0.2cm}
\noindent \textbf{Conformal map to the cylinder}
\vspace{0.2cm}

\begin{figure}
    \centering
    \tdplotsetmaincoords{10}{0}
    \begin{tikzpicture}[scale=1,tdplot_main_coords]
    \tdplotsetrotatedcoords{0}{10}{0}
    
    \begin{scope}[tdplot_rotated_coords]

    \begin{scope}[canvas is xy plane at z=0]
    \draw [smooth,domain=0:360,black] plot ({1.5*cos(\x)}, {1.5*sin(\x)});
    \end{scope}

    \begin{scope}[canvas is yz plane at x=0]
    \draw [smooth,domain=180:360,black] plot ({1.5*cos(\x)}, {1.5*sin(\x)});
    \draw [smooth,domain=0:180,black,dashed] plot ({1.5*cos(\x)}, {1.5*sin(\x)});
    \end{scope}

    \begin{scope}[canvas is xz plane at y=0]
    \draw [smooth,domain=180:360,black] plot ({1.5*cos(\x)}, {1.5*sin(\x)});
    \draw [smooth,domain=0:180,black,dashed] plot ({1.5*cos(\x)}, {1.5*sin(\x)});
    \end{scope}

    \draw[thick] (-6,-1.5,-6) -- (6,-1.5,-6) -- (6,-1.5,6) -- (-6,-1.5,6) -- cycle;
    
    \begin{scope}[canvas is xz plane at y=-1.5]
    \draw [smooth,domain=0:360,black,fill=gray] plot ({1.73*cos(\x)-3.46}, {1.73*sin(\x)});
    \end{scope}

    \begin{scope}[canvas is xz plane at y=-1.5]
    \draw [smooth,domain=0:360,black,fill=gray] plot ({1.73*cos(\x)+3.46}, {1.73*sin(\x)});
    \end{scope}

    \draw (0,1.5,0) -- ({-1.73-3.46},-1.5,0);
    \draw (0,1.5,0) -- ({1.73-3.46},-1.5,0);

    \begin{scope}[canvas is yz plane at x=-1.3]
    \draw [smooth,domain=0:180,black,dashed] plot ({0.75*cos(\x)}, {0.75*sin(\x)});
    \end{scope}

    \begin{scope}[canvas is yz plane at x=1.3]
    \draw [smooth,domain=0:180,black,dashed] plot ({0.75*cos(\x)}, {0.75*sin(\x)});
    \end{scope}

    \foreach \i in {1,...,30}
    {
    \begin{scope}[canvas is yz plane at x={-1.5*cos(\i)}]
    \draw [smooth,domain=0:360,gray,opacity=0.4] plot ({1.5*sin(\i)*cos(\x)}, {1.5*sin(\i)*sin(\x)});
    \end{scope}
    }

    \foreach \i in {1,...,30}
    {
    \begin{scope}[canvas is yz plane at x={1.5*cos(\i)}]
    \draw [smooth,domain=0:360,gray,opacity=0.4] plot ({1.5*sin(\i)*cos(\x)}, {1.5*sin(\i)*sin(\x)});
    \end{scope}
    }

    \begin{scope}[canvas is yz plane at x=-1.3]
    \draw [smooth,domain=180:360,black] plot ({0.75*cos(\x)}, {0.75*sin(\x)});
    \end{scope}

    \begin{scope}[canvas is yz plane at x=1.3]
    \draw [smooth,domain=180:360,black] plot ({0.75*cos(\x)}, {0.75*sin(\x)});
    \end{scope}

    \draw (0,1.5,0) -- ({1.73+3.46},-1.5,0);
    \draw (0,1.5,0) -- ({-1.73+3.46},-1.5,0);

    \draw[dashed] (-1.5,-3,-6) -- (-1.5,-3,6) -- (-1.5,3,6) -- (-1.5,3,-6) -- cycle;
    
    \end{scope}
    \end{tikzpicture}
    \caption{The stereographic mapping from the plane with two disks removed to the sphere is conformal, and takes the twice punctured plane to a band around the sphere. A second stereographic map from the sphere to the dashed plane takes the band to an annulus. The radial conformal map then maps the annulus to the cylinder.}
    \label{fig:stereographic}
\end{figure}
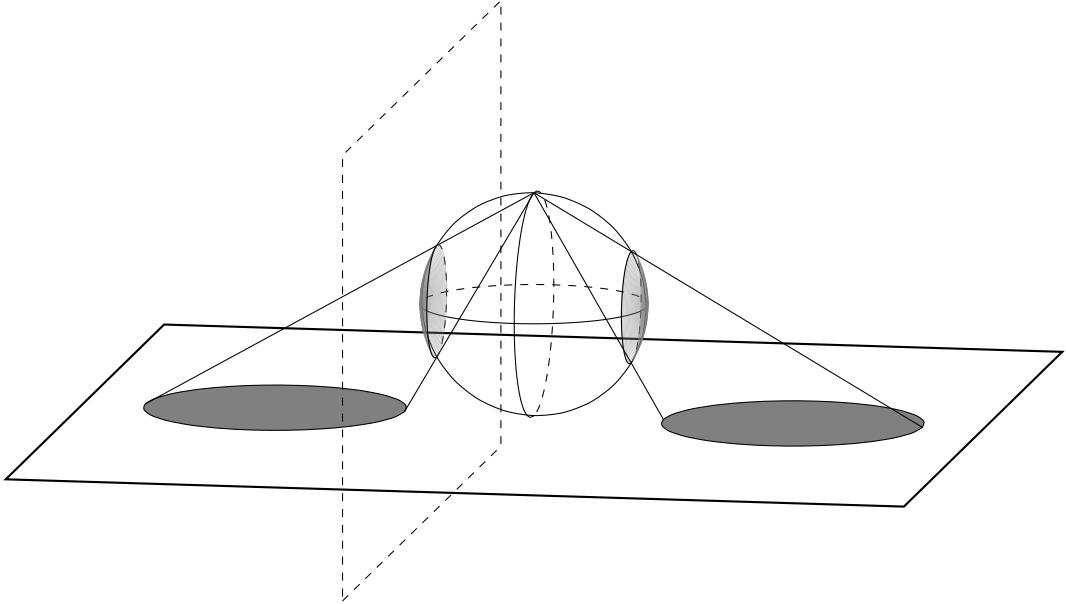

To map our plane with two disks removed to the cylinder, we will first map to the stereographic sphere, then project back down into a rotated plane. 
In the rotated plane, the resulting region is an annulus, which is conformal to a cylinder with finite height by the usual exponential map.
The basic idea is shown in figure \ref{fig:stereographic}.

To go from the stereographic coordinates to Cartesian coordinates, we have the map
\begin{align}
    x &= \frac{2 r \cos \phi }{1 - \sin \phi \cos\theta}, \nonumber \\
    \tau &= \frac{2 r \sin \phi \sin \theta}{1 - \sin \phi \cos\theta}.
\end{align}
We can check that constant $\phi=\phi_0$ curves map to circles in the $(x,\tau)$ plane. 
The circle parameters $(x_0,R)$ are related to $(r,\phi_0)$ by
\begin{align}
    R &= 2 r \tan \phi_0, \nonumber \\
    x_0 &= 2r \sec \phi_0.
\end{align}
Recall that to make the Wick rotation, we needed to set $x_0^2-R^2=1$. 
Here this condition amounts to $r=1/2$. 
We can also invert the above (keeping $r$ free), to find
\begin{align}
    r &= \frac{1}{2}\sqrt{x_0^2-R^2}, \nonumber \\
    \tan \phi_0 &= \frac{R}{\sqrt{x_0^2-R^2}}.
\end{align}

Next, we will map to a second plane, this one tilted 90 degrees compared to the first.
We show the setting again in figure \ref{fig:stereographic}. 
The new plane coordinates $\bar{x},\bar{\tau}$ are given in terms of the spherical coordinates by
\begin{align}\label{eq:phitoxbar}
    \bar{x} &= \frac{2r}{1-\cos\phi} \sin \phi \cos\theta, \nonumber \\
    \bar{\tau} &= \frac{2r}{1-\cos\phi} \sin \phi \sin\theta.
\end{align}
This takes the two caps of the sphere to a disk and an punctured plane, so that the remaining CFT lives on an annulus. 
The inner and outer radii are
\begin{align}
    r_+ &= \frac{2r \sin \phi_0}{1 - \cos \phi_0} = \frac{R \sqrt{x_0^2 - R^2}}{x_0-\sqrt{x_0^2-R^2}}, \nonumber \\
    r_- &= \frac{2r \sin \phi_0}{1 + \cos \phi_0} = \frac{R \sqrt{x_0^2 - R^2}}{x_0+\sqrt{x_0^2-R^2}}.
\end{align}
We can also invert \ref{eq:phitoxbar} to obtain
\begin{align}
    \tan \theta &= \frac{\bar{\tau}}{\bar{x}} \nonumber \\
    \sin \phi &= \frac{4r \sqrt{\bar{x}^2+\bar{\tau}^2}}{\bar{x}^2+\bar{\tau}^2 + 4r^2}
\end{align}

The coordinate change from the initial $(x,\tau)$ plane to the $(\bar{x},\bar{\tau})$ plane is
\begin{align}
    \bar{x} &= 2r \frac{x^2 + \tau^2 -4r^2}{(x-2r)^2 + \tau^2} \nonumber \\
    \bar{\tau} &= 8r^2\frac{\tau}{(x-2r)^2+\tau^2}
\end{align}
We can check explicitly that this is conformal, and maps circles centered around the origin in one set of coordinates to circles offset in the $x$ direction in the other coordinates.

Finally, we go to the cylinder. 
Using radial coordinates $(r,\phi)$ in the plane, the map to the cylinder is given by $r=e^X$. 
The height of the cylinder is therefore
\begin{align}\label{eq:cylinderheight}
    H = X_+ - X_- = \ln \left( \frac{r_+}{r_-}\right) = \ln\left( \frac{x_0 + \sqrt{x_0^2-R^2}}{x_0 - \sqrt{x_0^2-R^2}} \right).
\end{align}

\vspace{0.2cm}
\noindent \textbf{Comparison to the connected solution}
\vspace{0.2cm}

Recall from section \ref{appendix:BTZ} equation \ref{eq:X0conditionforaction} that the finite cylinder is in the disconnected phase when
\begin{align}
    H \geq \pi
\end{align}
which using \ref{eq:cylinderheight} leads to
\begin{align}
    1\leq x_0 \leq \frac{e^\pi+1}{e^{\pi}-1}.
\end{align}
Note the first inequality is automatically satisfied by our previous condition for Wick rotation, $x_0^2-R^2=1$. In the next section, we will relate the shift parameter $x_0$ to the angular opening of the brane when viewed in global AdS.
In particular, we find $x_0=\sec \Delta \phi$ for $\Delta \phi$ the angular radius of the region. 
The condition on $\Delta \phi$ then is $\Delta \phi \lesssim 24^\circ$.

\subsection{Disconnected branes in global AdS\texorpdfstring{$_{2+1}$}{TEXT}}

The disconnected solution is
\begin{align}
    (x \pm \ell x_0)^2 + \tau^2 + z^2 = R^2
\end{align}
Now we use the coordinate transformation \ref{eq:globaltoPoincareEuclidean} again to express this solution in global coordinates, finding 
\begin{align}
    \frac{1}{2}(e^{\tau_G} + e^{-\tau_G}(x_0^2-R^2)) = \mp x_0\sin(r) \cos(\phi)
\end{align}
To obtain Lorentzian global AdS, we will Wick rotate $\tau_G \rightarrow it_G$. 
To ensure this brane solution is well defined after the Wick rotation, we need to set $x_0^2-R^2=1$, so that the surface remains real. 
Doing so and Wick rotating, we obtain
\begin{align}
    \cos(t_G) = \mp x_0 \sin(r) \cos(\phi)
\end{align}
We can identify the two choices of sign with a shift in the angular coordinate, so that we have two branes in the geometry, with
\begin{align}
    \cos(t_G) &= x_0 \sin(r) \cos(\phi) \nonumber \\
    \cos(t_G) &= x_0 \sin(r) \cos(\phi-\pi)
\end{align}
Finally, we can identify $x_0$ with $1/\cos(\Delta \phi)$ by looking at this equation restricted to $r=\pi/2$, $t=0$. 

\subsection{Minimal surfaces}

Next we need to find minimal surfaces in the global geometry. 
We consider an interval ending at $t_G$, and $\phi=0,\pi$, and look for the minimal surface that encloses it.

The transformation from global to Poincare, in Lorentzian signature, is
\begin{align}\label{eq:loretziantransformations}
    t &= \frac{\ell \sin(t_G)}{\cos(t_G)-\sin(\phi)\sin(r)} \nonumber \\
    x &= \frac{\ell \cos(\phi)\sin(r)}{\cos(t_G)-\sin(\phi)\sin(r)} \nonumber \\
    z &= \frac{\ell \cos(r)}{\cos(t_G)-\sin(\phi)\sin(r)}
\end{align}
Under this coordinate transformation, we get to the global AdS metric \ref{eq:pureAdSglobal} with the brane
\begin{align}
    \cos(t_G) = \frac{\ell^2+R^2}{\ell^2-R^2}\sin(\phi) \sin(r).
\end{align}
Notice that $R$ is related to $\Delta \phi$ differently than when we transformed from Euclidean Poincare to Euclidean global coordinates, in particular
\begin{align}\label{eq:epsilonDeltaPhi}
    \frac{\ell^2+R^2}{\ell^2-R^2} = \sec(\Delta \phi).
\end{align}

We are interested in minimal surfaces anchored to $\phi=\pm \pi/2$, $t_G=s$. 
This means $t=\tan(s)$, $x=\pm \sec(s)$, so that, using equation \ref{eq:poincareareaformula} for areas of geodesics in Poincare, the brane anchored surface has area
\begin{align}
    A_b = \ell \log \left(\frac{\ell^2 -R^2}{R \delta} \right)
\end{align}
where the cutoff is at $z=\delta$. 
To compare this to the connected solution, we will need to translate this to a $r$ cut-off. 
Using \ref{eq:loretziantransformations} with $\phi=0$, we find
\begin{align}
    \delta = \frac{\epsilon}{ \cos(t_G)+ 1}
\end{align}
so that the disconnected minimal surface has area
\begin{align}
    A_b = 2\ell \log \left(\frac{\cos(t)+1}{\epsilon \tan \Delta \phi} \right)
\end{align}
where we used \ref{eq:epsilonDeltaPhi} to replace $R$ in favor of $\Delta \phi$. 
The minimal surface extending directly though the bulk at constant $t$, $\phi$, has area 
\begin{align}
    A_0 = 2\ell \log \left( \frac{2}{\epsilon} \right).
\end{align}
Requiring $A_0\leq 2 A_b$ we have
\begin{align}
    \boxed{\cos(t) \geq 2\tan\Delta \phi - 1}
\end{align}
Recall that in our protocol, we placed the cut at the top of the diamonds $V_i$, which will be at time $t = \pi/2-\Delta \phi$. 
Inserting this into the condition above, we find that we need $\Delta \phi \leq 39^\circ$ for the connected surface to be minimal. 

\bibliographystyle{JHEP}
\bibliography{biblio.bib}

\end{document}